\documentclass{classarxiv}
\usepackage{cite}
\usepackage{amsmath,amssymb,amsfonts}
\usepackage{algpseudocode}
\usepackage{algorithm}
\usepackage{graphicx,color}
\usepackage{textcomp}
\usepackage{soul,xcolor}
\sethlcolor{yellow}
\def\BibTeX{{\rm B\kern-.05em{\sc i\kern-.025em b}\kern-.08em
    T\kern-.1667em\lower.7ex\hbox{E}\kern-.125emX}}

\begin{document}

\title{Design of an embedded hardware platform for cell-level diagnostics in commercial battery modules}

\author{GABRIELE~MARINI$^\text{1,b}$\thanks{G. Marini and A. Colombo contributed equally to this work.},
ALESSANDRO~COLOMBO$^\text{2,b}$\thanks{G. Marini and A. Colombo contributed equally to this work.},
ANDREA~LANUBILE$^\text{1}$,
WILLIAM~A.~PAXTON$^\text{3}$,
and SIMONA~ONORI$^\text{1,a}$, FELLOW, IEEE
}

\affil{G. Marini, A. Lanubile, and S. Onori are with the Department of Energy Science and Engineering, Stanford University, Stanford, CA 94305 USA (e-mail: marinig@stanford.edu; lanubile@stanford.edu; sonori@stanford.edu).}

\affil{A. Colombo is with the Dipartimento di Elettronica, Informazione e Bioingegneria (DEIB), Politecnico di Milano, 20133 Milan, Italy (e-mail: alessandro4.colombo@polimi.it). He was with the Department of Energy Science and Engineering, Stanford University, Stanford, CA 94305 USA, at the time of this research.}

\affil{W. A. Paxton is with the Innovation and Engineering Center California, Volkswagen Group of America Inc., Belmont, CA 94002 USA (e-mail: William.Paxton@vw.com).}

\authornote{$^\text{b}$G. Marini and A. Colombo contributed equally to this work. \\ This work is supported by VW Group of America, Inc.}

\corresp{$^\text{a}$CORRESPONDING AUTHOR: Simona~Onori (e-mail: sonori@stanford.edu).}

%\markboth{Preparation of Papers for IEEE OPEN JOURNALS}{Marini \textit{et al.}}

\begin{abstract}
While battery aging is commonly studied at the cell-level, evaluating aging and performance within battery modules remains a critical challenge. Testing cells within fully assembled modules requires hardware solutions to access cell-level information without compromising module integrity. In this paper, we design and develop a hardware testing platform to monitor and control the internal cells of battery modules contained in the Audi e-tron battery pack. The testing is performed across all 36 modules of the pack. The platform integrates voltage sensors, balancing circuitry, and a micro-controller to enable safe, simultaneous cell screening without disassembling the modules. Using the proposed testing platform, cell voltage imbalances within each module are constrained to a defined reference value, and cell signals can be safely accessed, enabling accurate and non-invasive cell-level state-of-health assessments. On a broader scale, our solution allows for the quantification of internal heterogeneity within modules, providing valuable insights for both first- and second-life applications and supporting efficient battery pack maintenance and repurposing.
\end{abstract}


%\IEEEspecialpapernotice{(Invited Paper)}

\maketitle

\section{INTRODUCTION}  \label{Introduction}
\IEEEPARstart{T}{he} electric vehicle (EV) market has experienced exponential growth in recent years, and this trend is likely projected to continue as EVs play a crucial role in global decarbonization efforts \cite{muratori2025trends}. This growth has driven significant advancements in lithium-ion batteries (LIBs), which are renowned for their long cycle life, high energy density, and low self-discharge rates \cite{bibra2022global}. These characteristics have made LIBs indispensable across various sectors other than transportation, such as consumer electronics, medical equipment, power tools, renewable energy storage systems and data centers \cite{LiIonApp}. 

LIB packs are arranged in modules, each comprising multiple cells connected electrically and thermally, and meeting high energy requirements requires a large population of interconnected components. Consequently, small cell-to-cell electrochemical differences can translate into non-homogeneous behavior (e.g., uneven current sharing among parallel-connected cells and voltage imbalance among series-connected cells) and can propagate from cell to module and pack, ultimately affecting pack performance and lifetime.  For these reasons, internal pack screening and diagnostics are essential for assessing variability and ensuring reliable operation. Such screening can be performed either at the cell level or at the module/pack level \cite{pesaran1997thermal,mod_test}. 

Heterogeneity characterization becomes even more critical for battery repurposing, i.e., when EV battery packs approaching the end of their service life are reused in less demanding second-life applications (e.g., commercial/industrial energy storage systems) \cite{GU2024,moy2024second,bach2024fair,zhuang2025technoeconomic}, where identifying underperforming and close to failure components is key to enabling reliable battery reuse.

In the remainder of the paper, we adopt the $x$P$y$S notation to indicate the module/pack topology, where $x$ denotes the number of cells connected in parallel (to form a parallel array) and $y$ denotes the number of such parallel arrays connected in series. When either $x$ or $y$ equals 1, it is omitted.

The open literature reports module-level testing conducted either on laboratory-built modules or on commercially available systems. In the case of lab-built modules, small, low-voltage configurations (up to 5 V) composed of parallel-connected cells have attracted increasing attention.  For example, \cite{TEMP1Fill} investigates the effects of thermal gradients on current distribution, while \cite{naylor2024degradation} analyzes their impact on cell degradation. In addition, \cite{TEMP3LI} compares different module topologies to evaluate the effectiveness of air ventilation, and \cite{thermal} examines thermal runaway propagation in parallel-connected cells.
A number of studies focus on current imbalances in parallel configurations driven by multiple factors. 
In \cite{C2C1DIAO}, current distribution is examined between LiFePO$_4$ and Li(NiCoAl)O$_2$ cells, while \cite{TEMP2JOCHER} proposes mitigating these imbalances through the introduction of a controllable interconnection resistance. Similarly, \cite{C2C2TIAN} models the impact of cell inconsistencies on current distribution and validates the approach on aged cells, whereas \cite{TOP1GRUN} investigates the influence of current-collector geometry and resistance.  In \cite{weng2024current}, current imbalances and their impact on degradation are analyzed in a 2P array.  \cite{PIOMBO2024110783} investigates capacity and resistance heterogeneities in a 4P module with Hall-sensor-based current measurements (dataset in \cite{PIOMBO2024110227}), while \cite{wong2024differential} uses Differential Voltage Analysis (DVA) to reveal imbalance effects on differential-voltage features. 
For series-connected cells, \cite{thermal} investigates thermal-runaway propagation in a 9S module configuration, and \cite{allam2016characterization} studies aging propagation in a 4S string. \cite{preger2025impact} introduces a fully instrumented, reconfigurable module (up to 8S/8P) to track current, voltage, and energy distributions over aging. Finally, \cite{offer2012module} develops a 504-cell pack for a prototype EV and uses pack-level data for fault detection. In contrast, academic studies on commercially available battery modules are limited, primarily due to the need for high-voltage testing equipment. In \cite{intro}, a screening method is proposed for a retired EV pack, although it does not resolve individual cell properties. 
\cite{WASSILIADIS2022100167} investigates modules from a Volkswagen ID.3 Pro Performance pack, where cell-level data are obtained through irreversible disassembly, followed by individual cell testing within the module casing to preserve mechanical and thermal conditions. Similarly, \cite{BAUMANN2018295} analyzes modules from a Mercedes-Benz Vito e-Cell pack via accelerated module-level aging and periodic cell-level reference tests, requiring repeated disassembly and reassembly.
\cite{seriesZhou} conducts accelerated aging tests at the module and pack levels using modules manufactured by Lishen Battery Company Ltd. to develop a state-of-health (SOH) estimator.  Overall, these studies rely on module disassembly to access cell-level signals. To date, the only openly reported example enabling non-destructive internal module access is the Nissan Leaf 2P2S module, which incorporates a third terminal providing direct electrical access to each 2P block \cite{partcap,gao2024evaluation,cui2024taking}.
To the best of the authors’ knowledge, a non-disruptive hardware platform for cell-level screening of series-connected cells within commercial modules is still unavailable. This work addresses this gap by introducing a custom-designed, low-power embedded system capable of non-invasively measuring individual cell voltages without opening the module casing. The platform also integrates cell balancing functionalities to enhance safety and reliability.  Building on these measurements, we propose a methodology to quantify cell heterogeneities in terms of capacity, energy, and resistance. The paper is organized as follows: Section II presents the experimental setup; Section III details the methodology for extracting cell-level metrics; Section IV reports the analysis across all cells; and Section V concludes the paper.

\section{EXPERIMENTAL SETUP AND DESIGN}
This section is organized as follows: module configuration, battery cycling equipment, embedded monitoring and balancing circuitry, and testing profile design.

\begin{figure}[!b]\centering
	\includegraphics[width=0.87\linewidth]{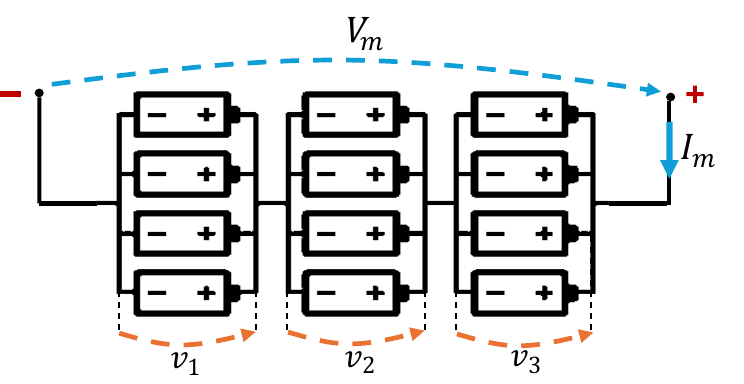}
	%\vspace{-1.0em}
    \caption{Schematic 4P3S topology; $v_1$, $v_2$, $v_3$ are the \textit{Cells} terminal voltage; $I_m$ module current; $V_m$ module voltage.}\label{fig:images_ppt_1}
\end{figure}

\subsection{MODULE CONFIGURATION}
\begin{figure*}[!h]\centering
	\includegraphics[width=1\linewidth]{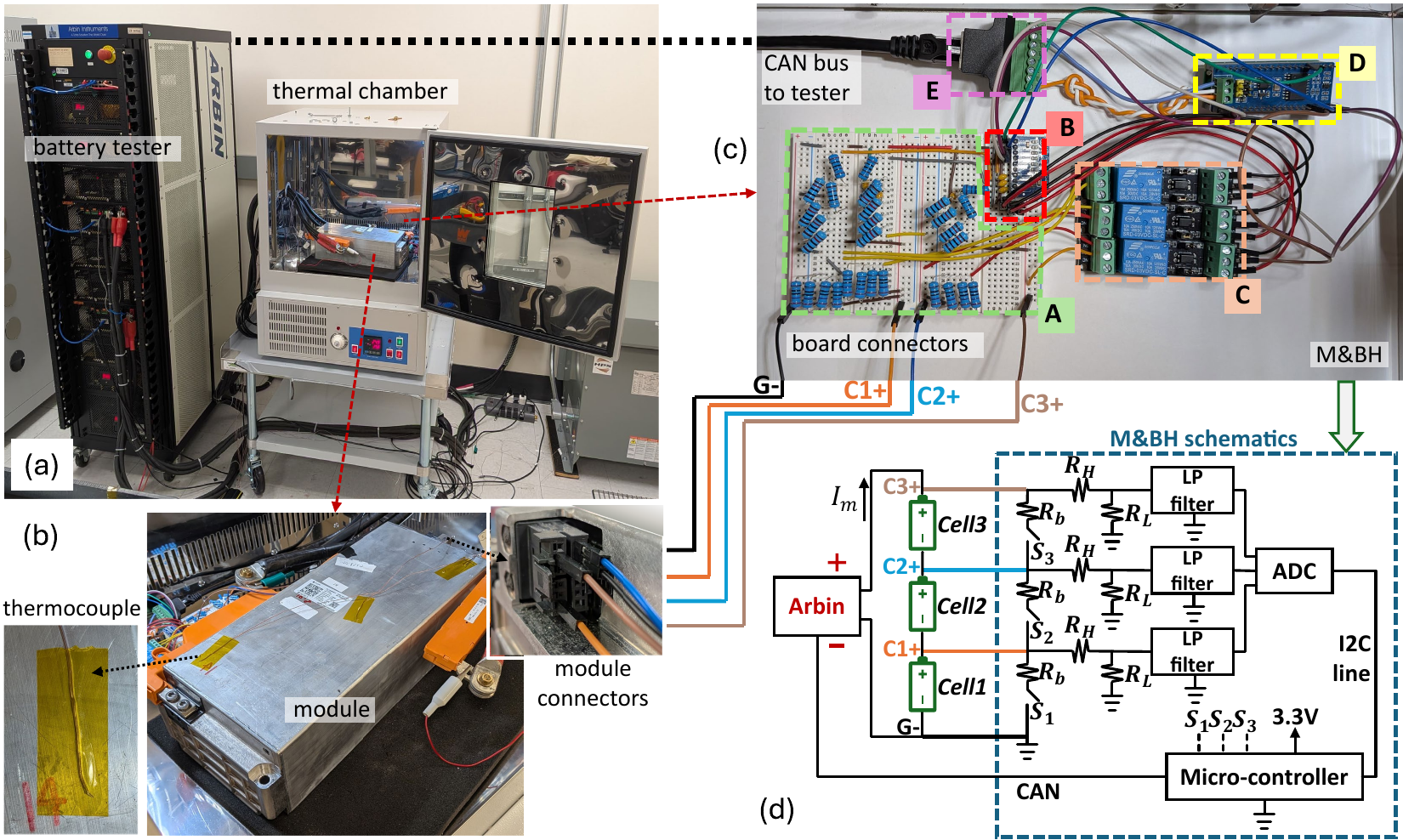}
	%\vspace{-2.0em}
    \caption{(a) Battery tester and thermal chamber with the module mounted inside; (b) detailed view of the module, including external connectors and thermocouple placement (three thermocouples attached to the module casing); (c) components of the M\&BH: (A) analog electrical circuit, (B) ADC, (C) balancing switches, (D) microcontroller, and (E) CAN adapter for the cycler; (d) schematic of the overall setup, including the battery tester, the module (series-connected \textit{Cells}), and the M\&BH. The module-to-board connections are highlighted with different colors: G$-$ (black) is the module ground,  C1$+$ (orange), C2$+$ (blue), and C3$+$ (brown) are the \textit{Cell} positive terminals.} \label{fig:photo_setup}
\end{figure*}

This study is based on modules from a 95 kWh battery pack from an all-electric Audi e-tron\textregistered. The vehicle was driven in the San Francisco Bay Area, CA, from September 2021 through January 2024, after which the pack was disassembled at the VW North America facility in Belmont, CA. Individual modules were extracted and then tested one by one in the Stanford Energy Control Laboratory, Stanford University \cite{catenaro2021experimental} from February through May 2024. The pack consists of 36 modules connected in series, each consisting of 12 NMC pouch cells arranged in a 4P3S configuration, as shown in Fig. \ref{fig:images_ppt_1}. 
In this work, each 4P group is represented as a single lumped element, denoted as \textit{Cell}, since the parallel-connected cells share the same terminal voltage and individual currents are not accessible. Each module comprises three \textit{Cells}, for a total of 108 \textit{Cells} in the pack. Each \textit{Cell} carries the module current $I_m$. 
Each module is equipped with external connectors (G-, C1+, C2+, and C3+, see Fig. \ref{fig:photo_setup}), whose potential differences provide the \textit{Cell} terminal voltages, denoted as  $v_j, \ j\in\{1,2,3\}$. Battery specifications in terms of rated voltage, capacity and energy are summarized in Table \ref{table:bspecs}. 
Physical cell specifications are taken from the datasheet, whereas \textit{Cell}, module, and pack specifications are derived from the electrical topology. The Cells operate within a voltage window of  $v_{min}=3.3$ V and $v_{max}=4.2$ V, respectively, which directly translates to a module voltage range of [$9.9$, $12.6$] V and pack voltage range of [$356.4$, $453.6$] V.
 
%\vspace{-2em}
\begin{table}[!h]
\centering
\caption{Battery cell, \textit{Cell}, module and pack specifications.}
\resizebox{1.00\columnwidth}{!}{%
\begin{tabular}{cccc}
\hline
                                                    & Rated        & Rated         & Rated        \\
                                                    & voltage {[}V{]}   & capacity {[}Ah{]} & energy {[}kWh{]} \\
\hline
cell                                                & 3.6   & 61.2  & 0.22   \\
\textit{Cell} (4P)                                       & 3.6   & 244.8 & 0.88   \\
module (4P3S)                                              & 10.8  & 244.8 & 2.64   \\
pack (4P108S)                                                & 388.8 & 244.8 & 95.04  \\
\hline
\end{tabular}}
\label{table:bspecs}
\end{table}

\subsection{BATTERY CYCLER AND THERMAL CHAMBER}
The laboratory setup for module testing is illustrated in Fig. \ref{fig:photo_setup} (a). An Arbin LBT22013 battery cycler is used for test actuation, with modules placed inside an Amerex IC150R thermal chamber to maintain a controlled ambient temperature. Surface temperature is monitored via three thermocouples attached to the module casing (Fig. \ref{fig:photo_setup}(b)). A safety cutoff is enforced to terminate the test if any thermocouple exceeds 50$^\circ$C. The three measurements are consistent (see Appendix III), and the module temperature is defined as their average.
The Arbin system records the module current and voltage.

\subsection{ELECTRONIC CIRCUIT DESIGN}
To monitor internal voltages without opening or altering the module casing, a dedicated electronic board, referred to as the Monitoring and Balancing Hardware (M\&BH), is developed. The hardware, shown in Fig. \ref{fig:photo_setup}(c) with schematic in Fig. \ref{fig:photo_setup}(d), comprises:
\begin{itemize}
\item[A)] a board featuring an analog electrical circuit composed of a resistor/capacitor (RC) network;
\item[B)] an analog-to-digital converter (ADC);
\item[C)] balancing switches;
\item[D)] a micro-controller;
\item[E)] a CAN adapter for communication to/from the cycler.
\end{itemize}
The parameter values for the analog circuit are reported in Appendix I (Table \ref{table:board_specs}). The tolerances on the values are $\pm$1\%. In the following, the M\&BH features are described in detail.

\subsubsection{MEASURING THE \textit{Cell} VOLTAGES} 
Voltage measurement is enabled by direct access to the \textit{Cell}-level potentials through the module external connectors (Fig. \ref{fig:photo_setup}(b)), which provide the positive terminal voltages (C1+, C2+, C3+) with respect to the common ground (G-).
These signals are first scaled using voltage dividers with resistive pairs 
$R_H$ and $R_L$, which scale the potentials by a factor of $\frac{R_L}{R_H+R_L}=0.25$ to 
 to match the ADC input range. The scaled voltages are then filtered by an analog low-pass RC filter with cutoff frequency  $f_{LP}=112.9$ Hz and subsequently digitized by an ADC (ADS1115) at a sampling rate of $10$ Hz.
This measurement pipeline follows standard BMS practice, where cell voltages are scaled through voltage dividers, filtered, and digitized \cite{niu2024research,kurkin2025battery}. The microcontroller (Raspberry Pi Pico) then rescales the digitized signals to their original voltage range, reconstructs the corresponding  \textit{Cell} terminal voltages, and transmits them to the battery cycler via CAN at $10$ Hz. Additional details on the voltage measurement pipeline are included in Appendix I. 
The proposed platform is tailored to the considered module configuration comprising three \textit{Cells}; however, it can be readily extended to modules with an arbitrary number of series-connected \textit{Cells}. The requirement for external connectors providing cell-level potentials is not a limitation, as such connectors are standard in modern battery modules for BMS-based cell-level monitoring.

\subsubsection{\textit{Cell} BALANCING} 
\textit{Cell} balancing is used to mitigate voltage imbalances among \textit{Cells}. If left unaddressed, such imbalances can lead to over or under charging, thermal gradients, and, ultimately, reduced performance and safety. The implemented balancing strategy is a standard passive balancing scheme \cite{lee2011comparison,azimi2022extending}, 
in which excess charge from higher voltage \textit{Cells} is dissipated through parallel branches containing bleeding resistors $R_b$. 
Each \textit{Cell} is associated with a controllable switch connected in series with its corresponding resistor. For the $j$-th \textit{Cell}, the switch state $S_j$  determines the balancing operation:
\begin{itemize}
    \item  $S_j=0$ (open): the parallel branch of the $j$-th \textit{Cell} is inactive, and the entire module current flows through the \textit{Cell};
    \item $S_j=1$ (closed): part of the module current is diverted through the bleeding resistor, dissipating excess energy and activating balancing.
\end{itemize}

Each switch is independently controlled by the micro-controller based on the measured cell voltages. Given the binary nature of the switch position $S_j$, a threshold-based bang-bang balancing strategy is implemented. Since the objective is to limit voltage imbalances within each module, the control law for the $j$-th \textit{Cell} is defined as:
\begin{equation}
	S_j(v_1,v_2,v_3) = 
	\begin{cases}
		1,\ \text{ if }\ v_j \ge \min{(v_1,v_2,v_3)} + v_{th} \\
		0,\ \text{ otherwise } 
	\end{cases},
    \label{eq:balancing_alg}
\end{equation}
where $v_{th}$ is a predefined voltage threshold and $v_j$ is the voltage of the 
$j$-th \textit{Cell}.  The balancing switch is activated, $S_j=1$, for those \textit{Cells} whose voltage exceeds the minimum voltage within the triplet by more than $v_{th}$.	
As a result, no more than two switches can be closed simultaneously. After a balancing interval, the maximum voltage imbalance $\Delta v_{max}$ among \textit{Cells} aims not to exceed $v_{th}$:
\begin{equation}
\Delta v_{max}=\max(v_1,v_2,v_3)-\min(v_1,v_2,v_3)\leq v_{th}.
\end{equation}
The effectiveness of the balancing action depends on both the duration of the balancing phase and the initial value of $\Delta v_{max}$. In particular, for larger initial imbalances, a short balancing duration may be insufficient to achieve the desired equalization. It is important to notice that voltage regulation indirectly enables control of each \textit{Cell}'s state-of-charge (SOC) since higher \textit{Cell} voltages correspond to higher SOC.  By closing the balancing switch of an overcharged \textit{Cell}, the current flowing through it is reduced, thereby lowering its charging rate. In this work,  $v_{th}$ is set to $2.5$ mV and the bleeding resistance is chosen as $R_b=67.5~\Omega$ consistent in order of magnitude with on-board values. The effect of the bleeding resistance on SOC dynamics is analyzed in Appendix II.

\begin{figure*}[!t]
\centering
	\includegraphics[width=1\linewidth]{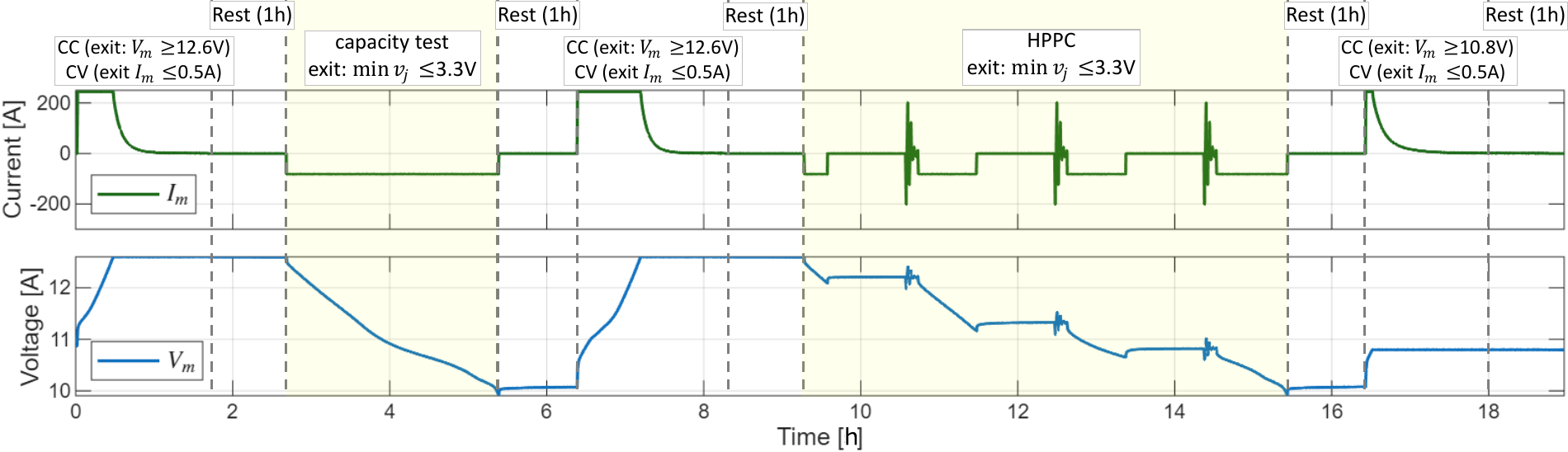}
	%\vspace{-2.0em}
    \caption{Top: actuated module current profile $I_m$. Bottom: resulting module voltage $V_m$ (for module no.29). The description and the exit conditions of each portion are reported; diagnostic segments are highlighted in yellow. By convention, positive current means charging; negative discharging.} \label{fig:current_profile_1}
\end{figure*}

\subsection{TESTING PROFILE}
Dedicated diagnostic tests were conducted to extract key aging-related metrics at the \textit{Cell} level. Specifically, discharge capacity and discharge energy are computed from capacity tests; the internal resistance using Hybrid Pulse Power Characterization (HPPC) tests. Fig. \ref{fig:current_profile_1} illustrates the designed module current profile, actuated using the Arbin cycler, along with the resulting module voltage. Additional details are provided below.

\subsubsection{CAPACITY TEST}
The capacity test consists of a C/3 discharge, corresponding to a constant current of 81.6 A. The applied C-rate is configurable, depending on the application and time constraints. Starting from 12.6 V, the module is discharged until the \textit{Cell} with the lowest voltage reaches the predefined lower threshold of 3.3 V (i.e., the cell cutoff voltage). This is followed by a one-hour rest period. Balancing is intentionally deactivated during this test as the resulting voltage imbalances at the end of discharge are required to quantify capacity heterogeneities among the \textit{Cells}, as described in Section \ref{sec:methodology}. The test is conducted at a controlled chamber temperature of 25$^{\circ}$C.

\subsubsection{HPPC}
The HPPC alternates C/3 discharges, one-hour resting intervals, and current pulses at different C-rates. It is conducted at three SOC levels, i.e. 90\%, 65\%, and 40\%. These SOC levels are reached via Coulomb counting during a C/3 discharge, using the rated  \textit{Cells} capacity as a reference (see, Table \ref{table:bspecs}). 
At each SOC level, a sequence of eight consecutive current pulses is applied, including  including 4/5C (200 A), C/2 (122.4 A), C/10 (24.48 A), and C/20 (12.24 A), each performed in both discharge and charge. Each pulse lasts 15 s and is followed by a 60 s rest period. At the end of the protocol, the module is discharged until the cell with the lowest voltage reaches 3.3 V, followed by a one-hour rest period. Balancing is disabled throughout the HPPC test. All tests are conducted at three controlled temperatures, namely, 15$^{\circ}$C, 25$^{\circ}$C and 35$^{\circ}$C.

\subsubsection{CC-CV CHARGING}
Before the diagnostic tests, the module is charged at 244.8 A (1C) constant current (CC) until its voltage reaches the upper cutoff value 12.6 V; after that, it is kept at constant voltage (CV) until the current decreases to 0.5 A (C/500). During the CC-CV phase, balancing is active, ensuring that all \textit{Cells} reach the same voltage (4.2 V) at the start of the diagnostic tests, with a maximum deviation of 2.5 mV according to the implemented balancing strategy.  
The final CC–CV charge is instead used to restore the cells to their rated voltage at the end of the testing profile. Each CC–CV charging phase is followed by a one-hour rest period. 

\begin{figure}[!b]
\centering
	\includegraphics[width=1\linewidth]{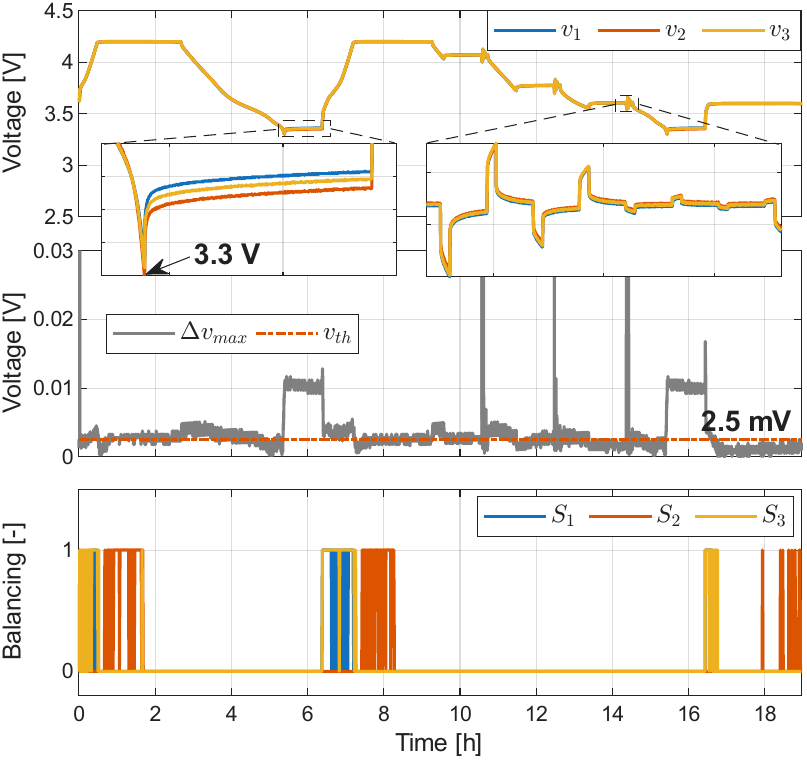}
	%\vspace{-2.0em}
    \caption{Top: measured \textit{Cell} voltages $v_1$, $v_2$, $v_3$, with zoomed-in views. Middle: maximum \textit{Cell} voltage imbalance $\Delta v_{max}$ and balancing threshold $v_{th}$. Bottom: switch states $S_1$, $S_2$, $S_3$ determined by \eqref{eq:balancing_alg}.}\label{fig;paper_plot_module}
\end{figure}

The resulting \textit{Cell}-level signals for module no.29 are shown in Fig. \ref{fig;paper_plot_module}. The balancing strategy described in \eqref{eq:balancing_alg}, active during the CC-CV phases,  effectively reduces voltage differences among the  \textit{Cells} and maintains them within the desired threshold.

\section{QUANTITATIVE METRICS COMPUTATION}\label{sec:methodology}
This section presents the methodology used to compute the metrics from the proposed experimental setup.

\subsection{DISCHARGE CAPACITY}
The discharge capacity of each \textit{Cell} within a module is obtained from the capacity test via Coulomb counting, i.e., by integrating the module current over a fixed voltage range. Starting from a common initial voltage (4.2 V $\pm$ 2.5 mV), the final voltages of the \textit{Cells} at the end of the test may differ due to capacity heterogeneities. In particular, the \textit{Cell} with the lowest capacity depletes its SOC faster and reaches the lower cutoff voltage first.
Quantifying discharge capacity over a common voltage range therefore provides a direct measure of capacity heterogeneity among the \textit{Cells}. A visualization of the resulting voltage imbalances at the end of the capacity test is shown in Fig.~\ref{fig;paper_plot_module} (zoom of the top plot, left).

Given the module current $I_m$ and a voltage window $[\underline{v}, \bar{v}]$, the discharge capacity $q_j$ of the $j$-th \textit{Cell} is computed as
\begin{equation}
    q_j=\frac{1}{3600}\int_{t_{0}:v_j(t_{0})=\bar{v}}^{t_{f}:v_j(t_{f})=\underline{v}} \left|I_m(t) \right|\mathrm{d}t .\label{eq:qj}
\end{equation}
The bounds $\bar{v}$ and $\underline{v}$ define the largest common voltage window across all \textit{Cells}, during the C/3 capacity test. 

Specifically, after testing all modules in the pack, $\bar{v}$ is set to the minimum of the initial voltages observed across all \textit{Cells}, while $\underline{v}$ is defined as the maximum final voltage. The iterative procedure used to compute $[\underline{v}, \bar{v}]$ is described in Algorithm \ref{alg:voltage_range}. This approach ensures a consistent \textit{Cell} capacity comparison not only within individual modules but also across the entire battery pack.
Algorithm \ref{alg:voltage_range} yields a slightly reduced voltage window, with $\underline{v}=3.3373$ V and $\bar{v}=4.1953$ V. The lower bound is about $37$ mV above the lower cutoff voltage (3.3 V) reflecting discharge imbalance driven by capacity heterogeneities. The upper bound is about $5$ mV below  $4.2$ V due to residual imbalances that persist despite the balancing algorithm, which may require longer balancing durations to fully converge within the $\pm 2.5$ mV threshold. In \eqref{eq:qj}, the integral is normalized by 3600 to convert the integrated current into ampere-hours (Ah).
Because all \textit{Cells} carry the same module current, the module discharge capacity $Q_m$ is limited by the \textit{Cell} with the lowest discharge capacity, namely,
\begin{equation}
    Q_m=\min (q_1, q_2, q_3).\label{eq:Qm}
\end{equation}
The \textit{Cell} with the lowest capacity is referred to as the \textit{weakest} \textit{Cell}. 
%Its index within each module is defined as
%\begin{equation}
 %   j^*=\arg\min(q_1, q_2, q_3). \label{eq:weakest_cell}
%\end{equation}

\begin{algorithm}[!h]
\caption{Voltage window [$\underline{v}$, $\bar{v}$] iterative computation}\label{alg:voltage_range}
\small
\begin{algorithmic}
\State $\bar{v}^{(\textbf{0})}\gets v_{max}$ \Comment{Initialize upper bound}
\State $\underline{v}^{(\textbf{0})} \gets v_{min}$ \Comment{Initialize lower bound}
\State $h=1$ \Comment{Initialize module counter}
\While{$h \leq N_{modules}=36$} \Comment{Iterate over modules}
    \State Execute testing profile on module $h$
    \State Extract the C/3 discharge voltage profiles $v_j(i)$, for $i=1,\dots,N_{samples}$ and for all $j=1,\dots,N_{\textit{Cells}}$
    \State $j=1$ \Comment{Initialize \textit{Cell} counter}
    \While {$j\leq N_{\textit{Cells}}=3$} \Comment{Iterate over \textit{Cells}}
        \State $\bar{v}^{(\textbf{h})} \gets \min{(\max_i{v_j(i)},\bar{v}^{(\textbf{h-1})})}$ \Comment{Update upper bound}
        \State $\underline{v}^{(\textbf{h})} \gets \max{(\min_i{v_j(i)},\underline{v}^{(\textbf{h-1})})}$ \Comment{Update lower bound}
        \State $j\gets j+1 $ \Comment{Increment \textit{Cell} counter}
    \EndWhile
    \State $h\gets h+1 $ \Comment{Increment module counter}
\EndWhile
\State $\bar{v}\gets \bar{v}^{(\textbf{h})}$ \Comment{Return upper bound}
\State $\underline{v}\gets \underline{v}^{(\textbf{h})}$ \Comment{Return lower bound}
\end{algorithmic}
\end{algorithm}

\subsection{DISCHARGE ENERGY}
Similarly, the discharge energy $e_j$ of the $j$-th \textit{Cell} test is obtained by integrating the power, i.e., $v_j(t) I_m(t)$, over the same voltage window:
\begin{equation}
    e_j=\frac{1}{3600}\int_{t_{0}:v_j(t_{0})=\bar{v}}^{t_{f}:v_j(t_{f})=\underline{v}} \left|v_j(t)I_m(t) \right|\mathrm{d}t .\label{eq:ej}
\end{equation}
The voltage window is the one defined by Algorithm \ref{alg:voltage_range}, and the integral is normalized by 3600 to express the discharge energy in watt-hours (Wh).

Since the module voltage is the sum of the \textit{Cell} voltages, the module discharge energy $E_m$ is given by
\begin{equation}
E_m=\sum_{j\in{1,2,3}} e_j. \label{eq:Em}
\end{equation}

The metrics in \eqref{eq:ej} and \eqref{eq:Em} include both the \textit{Cells}’ energy and the energy dissipated in the interconnect resistances due to the voltage measurement; nonetheless, the latter contribution is assumed negligible.

\begin{figure*}[!h]
\centering
\includegraphics[width=1\linewidth]{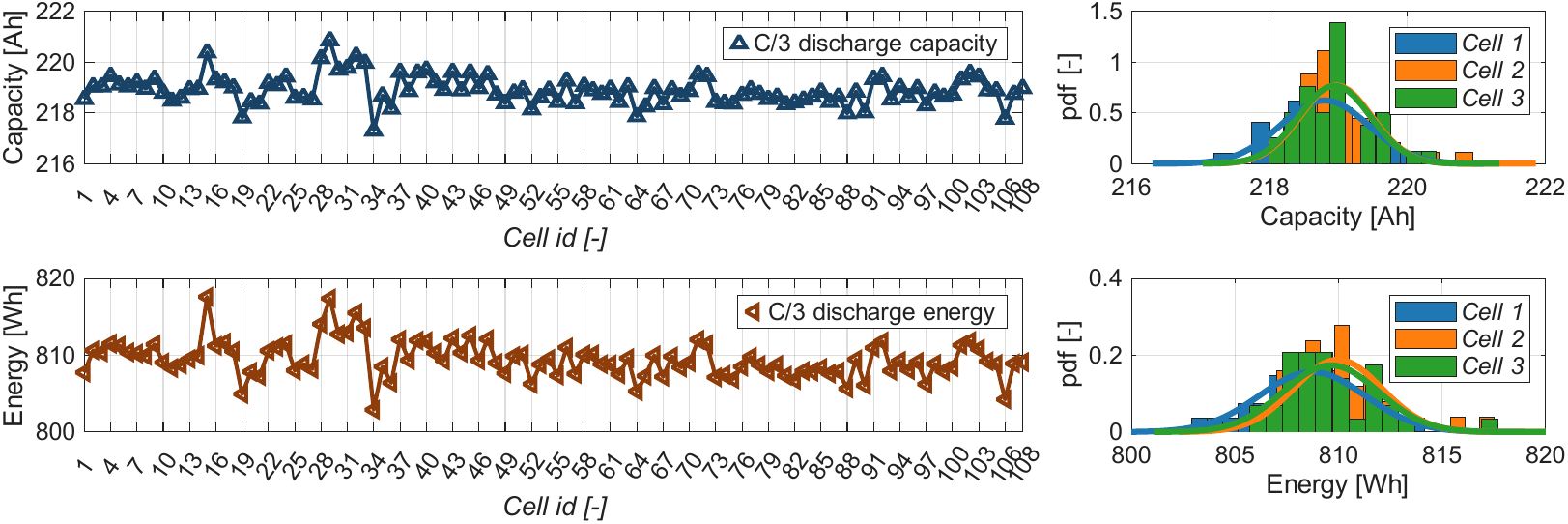}
%\vspace{-2.0em}
\caption{Top: C/3 discharge capacities for all 108 \textit{Cells} in the pack. Bottom: corresponding C/3 discharge energies. Right: capacity and energy distributions by \textit{Cell} position within each module, with Gaussian fits.}\label{fig:paper_cell_capacities_energies}
\end{figure*}

\subsection{INTERNAL RESISTANCE}
The internal resistance of each \textit{Cells} is estimated from the ohmic voltage drop induced by current pulses during the HPPC test. For each  SOC level $SOC\in\{40\%,65\%,90\%\}$ and temperature $T\in\{15^\circ C,25^\circ C,35^\circ C\}$ (i.e., nine SOC--temperature combinations), 
the resistance  $r_j$ of the $j$-th \textit{Cell} is computed as the ratio between the instantaneous voltage drop and the corresponding current pulse amplitude, under the assumption of zero current prior to the pulse (by design).
To improve robustness to noise, the resistance is averaged over all pulses applied at the same SOC level:
\begin{equation}
    r_j = \frac{1000}{N_{pulses}}\sum_{h=1}^{N_{pulses}}\left|\frac{\Delta v_{j,h}}{I_{pulse,h}}\right|,\label{eq:rj}
\end{equation}
where $N_{pulses}=8$, $I_{pulse,h}$ is the amplitude of the $h$-th pulse, and $\Delta v_{j,h}$ is the corresponding voltage drop. The factor 1000 converts the resistance to milli-ohms.  
A visualization of the ohmic voltage drops during the HPPC test is shown in Fig. \ref{fig;paper_plot_module}  (top panel, zoomed view). 

Internal resistance is an inherently instantaneous (pulse-based) metric and is therefore sensitive to acquisition latency and signal misalignment.  In our setup (see Appendix I), the board-to-Arbin communication introduces an approximately 100 ms delay between the module current signal (measured directly by the Arbin) and the Cell voltage signals (measured by the electronic board and transmitted via CAN). As a result, the onset of the current pulse does not exactly coincide with the corresponding voltage drop (see Fig.~\ref{fig:Latency}).
This misalignment is explicitly accounted for when computing  $\Delta v_{j,h}$ in \eqref{eq:rj}. Specifically, for each current pulse $h$, the voltage drop $\Delta v_{j,h}$ is evaluated by detecting the transition and computing the difference between the first voltage sample after the drop and the last sample before the drop.

Once the resistance values are computed for each SOC and temperature, a parabolic relationship between resistance $r$ and temperature $T$ is fitted
 to capture the observed inverse dependence between the two variables \cite{barcellona2022aging}. The model takes the following form:
\begin{equation}
    r(T,SOC)=\frac{a_1(SOC)}{T-a_2(SOC)}+a_3(SOC),\label{eq:r_T_relation}
\end{equation}
where the coefficients $\{a_m\}_{m=1,2,3}$ are SOC-dependent parameters.

\section{EXPERIMENTAL RESULTS} \label{results}
This section presents the main results obtained from 36 modules tested. These results provide a comprehensive characterization of all the \textit{Cells} of the pack and enable the identification of heterogeneities both within individual modules and across the entire pack.

\begin{table}[!b]
\centering
\caption{Discharge capacity and energy statistics based on the \textit{Cell} position within the module. Module-level stats are reported as well.}
\resizebox{0.95\columnwidth}{!}{%
\begin{tabular}{>{\centering\arraybackslash}p{1.5cm} >{\centering\arraybackslash}p{1cm} >{\centering\arraybackslash}p{0.90cm} >{\centering\arraybackslash}p{1.0cm} >{\centering\arraybackslash}p{1.0cm}}
\hline
\multicolumn{5}{c}{\textbf{C/3 discharge capacity}} \\
\hline
\textbf{} & mean [Ah] & std [Ah] & min [Ah] & max [Ah] \\
\hline
\textit{Cell} 1        & 218.80 & 0.64 & 217.31 & 220.16 \\
\textit{Cell} 2        & 218.96 & 0.52 & 218.26 & 220.86 \\
\textit{Cell} 3        & 218.95 & 0.52 & 218.03 & 220.39 \\
module-level  & 218.58 & 0.49 & 217.31 & 219.78 \\
\hline
\multicolumn{5}{c}{\textbf{C/3 discharge energy}} \\
\hline
\textbf{} & mean [Wh] & std [Wh] & min [Wh] & max [Wh] \\
\hline
\textit{Cell} 1        & 808.73  & 2.55 & 802.90 & 814.07 \\
\textit{Cell} 2        & 809.97  & 2.12 & 807.31 & 817.40 \\
\textit{Cell} 3        & 809.63  & 2.29 & 806.09 & 817.62 \\
module-level  & 2428.33 & 5.90 & 2417.84 & 2444.22 \\
\hline
\end{tabular}}
\label{table:cap_ene_stats}
\end{table}

\begin{figure}[!b]
\centering
\includegraphics[width=1\linewidth]{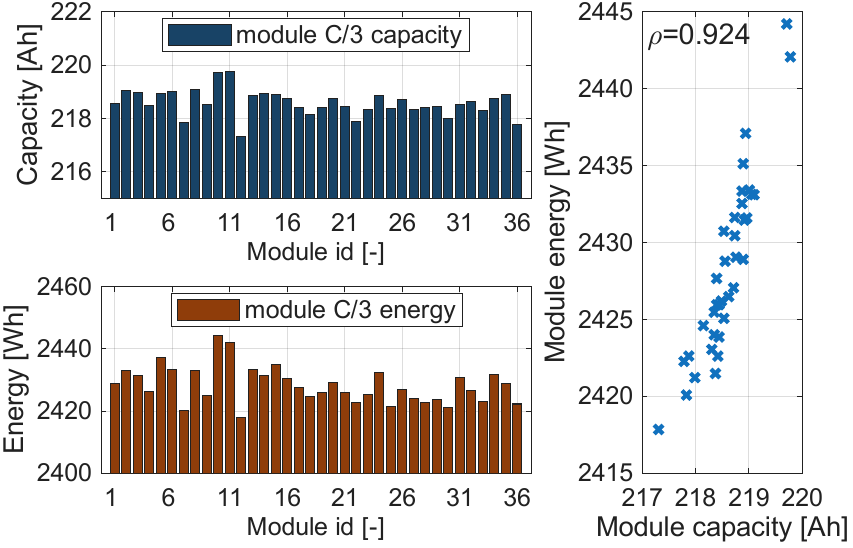}
%\vspace{-2.0em}
\caption{Left: module-level discharge capacity and energy. Right: correlation between capacity and energy.}\label{fig:paper_module_cap_ene}
\end{figure}

\begin{figure}[!b]
\centering
\includegraphics[width=1\linewidth]{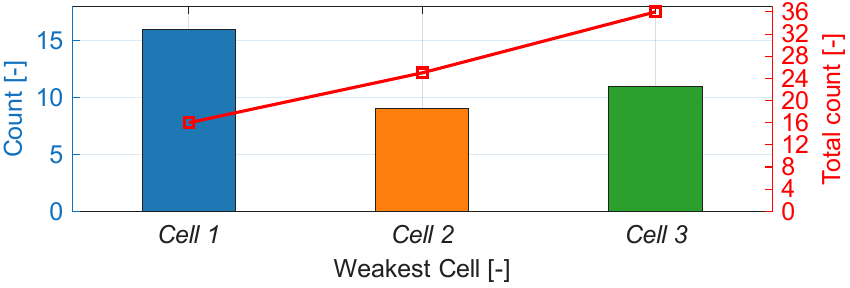}
%\vspace{-2.0em}
\caption{Histogram of the weakest \textit{Cells} across the 36 modules. Left axis: occurrences of the weakest \textit{Cell} across the three \textit{Cells} of the module. Right axis: cumulative count of occurrences (up to 36).}\label{fig:paper_cell_weakest}
\end{figure}

\begin{figure*}[!t]
\centering
\includegraphics[width=1\linewidth]{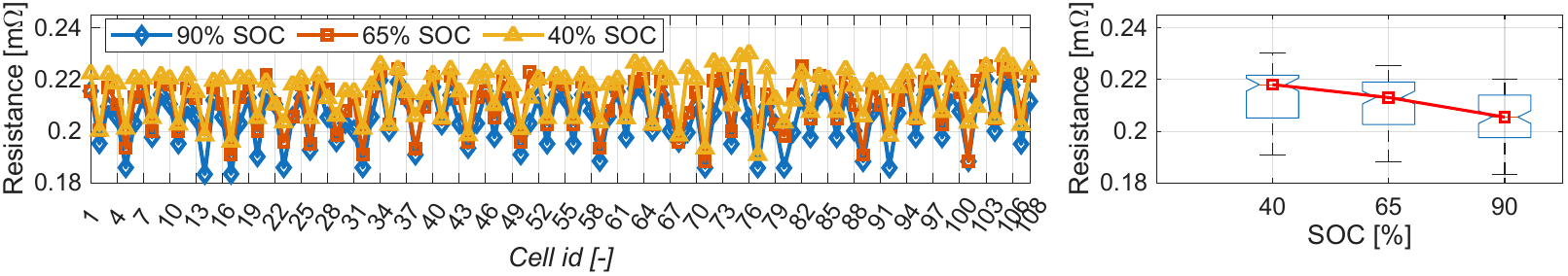}
%\vspace{-2.0em}
\caption{Left: resistances values for all 108 \ \textit{Cells} in the pack, computed at 25$^\circ$C, at  40\%, 65\% and 90\% SOC. Right: boxplots of the corresponding distributions; the red line connects the medians..}\label{fig:paper_reistances}
\end{figure*}

\subsection{CAPACITY AND ENERGY HETEROGENEITIES}
Fig. \ref{fig:paper_cell_capacities_energies} shows the C/3 discharge capacities (top) and C/3 discharge energies (bottom) for all \textit{Cells} in the pack, computed using \eqref{eq:qj}, \eqref{eq:ej} and Algorithm \ref{alg:voltage_range}. Each module test characterizes a set of three \textit{Cells}. 
On the left-hand side, capacities and energies are plotted according to the \textit{Cell} position within the pack (following the module ordering) highlighting intrinsic variations. The capacity and energy trends are consistent, with the lowest values observed for \textit{Cell} no. 34.
The spread of these values reflects the extent of heterogeneity across the pack.
On the right-hand side, the distributions of discharge capacity and energy are shown as a function of the \textit{Cell} position within each module. While the distributions are generally consistent across \textit{Cells}, \textit{Cell} 1 exhibits a slightly lower mean and higher variance compared to \textit{Cells} 2 and 3, indicating marginally different behavior (also supported by Gaussian fits).
The mean, standard deviation, minimum and maximum value of these distributions are listed in Table \ref{table:cap_ene_stats}.

The corresponding module-level capacity and energy values are shown in Fig. \ref{fig:paper_module_cap_ene}, computed according to \eqref{eq:Qm} and \eqref{eq:Em}. 
On the left-hand side, discharge capacity and energy for all modules are reported, with their statistical properties summarized in Table \ref{table:cap_ene_stats}. On the right-hand side, their correlation is shown, with a Pearson coefficient of $\rho=0.924$, indicating a strong relationship between the two quantities. Notably, module no.12 exhibits the lowest capacity and energy, consistent with the fact that it contains \textit{Cell} no.34, identified as the weakest \textit{Cell} in the pack.
A detailed look at the weakest \textit{Cell} within each module  is shown in Fig. \ref{fig:paper_cell_weakest}. 
\textit{Cell} 1 is identified as the weakest in 16 modules, while \textit{Cells} 2 and 3 are weakest in 9 and 11 modules, respectively. This uneven distribution suggests the presence of an intra-module gradient, consistent with the trends observed in Fig.~\ref{fig:paper_cell_capacities_energies}.
 %The non-uniformity highlighted by these findings should be considered in future assessments, particularly when evaluating the modules for repurposing applications. 

\subsection{RESISTANCE HETEROGENEITIES}
\begin{figure}[!b]
\centering
\includegraphics[width=1\linewidth]{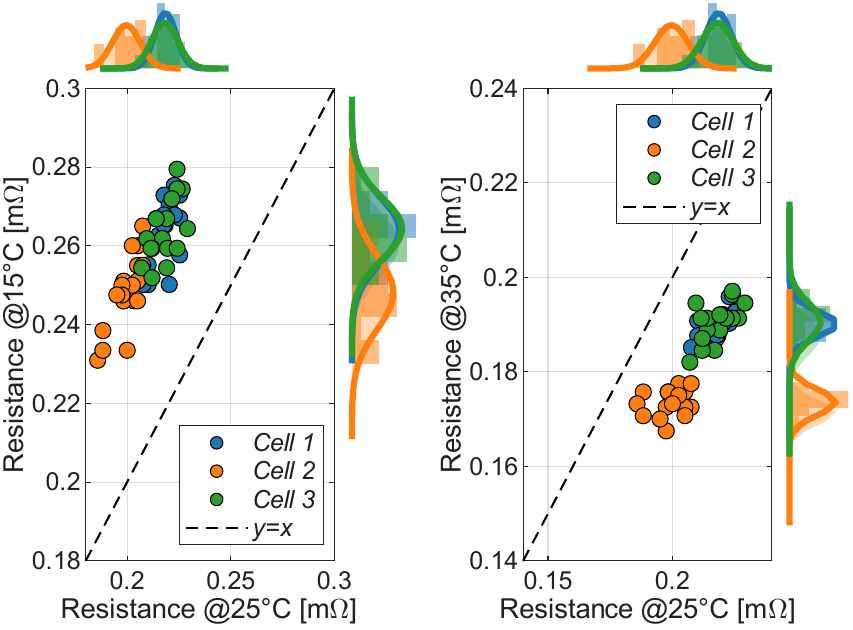}
%\vspace{-2.0em}
\caption{Comparison of resistance values at different setpoint temperatures as a function of \textit{Cell} position within the modules (color-coded). Left: 25\textdegree C vs. 15\textdegree C resistance; right: 25\textdegree C vs. 35\textdegree C resistance. Resistance distributions by \textit{Cell} position are also shown, with histograms fitted using Gaussian distributions..}\label{fig:paper_cell_reistances}
\end{figure}

The resistance values, computed according to \eqref{eq:rj}, are shown in Fig.~\ref{fig:paper_reistances} as a function of \textit{Cell} position, at 25$^\circ$C and across all SOC levels. While the distributions exhibit comparable variance, their medians differ, with resistance increasing at lower SOC. This indicates higher internal power losses under low-charge conditions, consistent with \cite{barcellona2022aging}.
Examining resistance as a function of \textit{Cell} position (left-hand side) reveals a clear periodic pattern, with local minima consistently occurring at the second \textit{Cell} within each module (i.e., the centrally positioned \textit{Cell}). This indicates structured intra-module resistance heterogeneity.
A more detailed analysis is provided in Fig.~\ref{fig:paper_cell_reistances}, which reports resistance as a function of \textit{Cell} position at the three setpoint temperatures, aggregating all SOC levels. The centrally positioned \textit{Cell} (position 2) forms a distinct cluster, while the outer \textit{Cells} (positions 1 and 3) largely overlap. This separation is confirmed by the mean values, with the central \textit{Cell} consistently exhibiting lower resistance across all temperatures (see Table~\ref{table:resistance_stats}). Additionally, resistance increases as temperature decreases and decreases as temperature increases.

\begin{figure}[!b]
\centering
\includegraphics[width=1\linewidth]{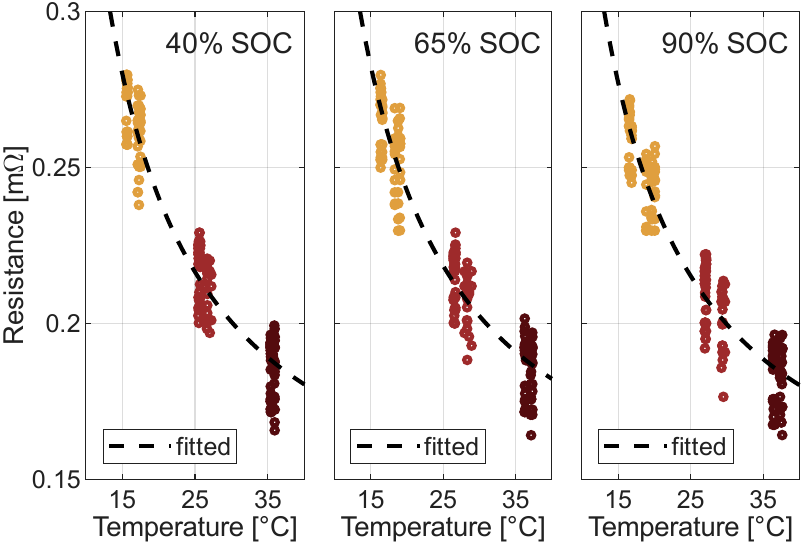}
%\vspace{-2.0em}
\caption{Scatter plot of the computed resistance values as a function of measured module temperature, across all SOC levels. The data are fitted with parabolic functions (dashed lines), according to \eqref{eq:r_T_relation}.}\label{fig:paper_reistances_temp}
\end{figure}

\begin{table}[!b]
\centering
\caption{Internal resistance statistics based on the \textit{Cell} position within the module, grouped by setpoint temperature.}
\resizebox{0.95\columnwidth}{!}{%
\begin{tabular}{>{\centering\arraybackslash}p{1.5cm} >{\centering\arraybackslash}p{1.0cm} >{\centering\arraybackslash}p{0.9cm} >{\centering\arraybackslash}p{1.0cm} >{\centering\arraybackslash}p{1.0cm}}
\hline
\multicolumn{5}{c}{\textbf{Internal resistance}} \\
\hline
\textbf{} & mean [m$\Omega$] & std [m$\Omega$] & min [m$\Omega$] & max [m$\Omega$] \\
\hline
\textbf{@15\textdegree C} & & & & \\
\textit{Cell} 1 & 0.2640  &  0.0083  &  0.2502  &  0.2754 \\
\textit{Cell} 2 & 0.2482  &  0.0095  &  0.2310  &  0.2650 \\
\textit{Cell} 3 & 0.2644  &  0.0078  &  0.2519  &  0.2795 \\
\hline
\textbf{@25\textdegree C} & & & & \\
\textit{Cell} 1 & 0.2185  &  0.0052  &  0.2077  &  0.2253 \\
\textit{Cell} 2 & 0.1996  &  0.0067  &  0.1858  &  0.2076 \\
\textit{Cell} 3 & 0.2181  &  0.0063  &  0.2070  &  0.2291 \\
\hline
\textbf{@35\textdegree C} & & & & \\
\textit{Cell} 1 & 0.1902  &  0.0027  &  0.1852  &  0.1959 \\
\textit{Cell} 2 & 0.1736  &  0.0027  &  0.1675  &  0.1775 \\
\textit{Cell} 3 & 0.1906  &  0.0041  &  0.1821  &  0.1971 \\
\hline
\end{tabular}}
\label{table:resistance_stats}
\end{table}

\begin{table}[!b]
\centering
\caption{Resistance vs temperature fitted model coefficients at the three tested SOC levels.}
\resizebox{0.95\columnwidth}{!}{%
\begin{tabular}{>{\centering\arraybackslash}p{1.5cm} >{\centering\arraybackslash}p{1.0cm} >{\centering\arraybackslash}p{0.9cm} >{\centering\arraybackslash}p{1.0cm} >{\centering\arraybackslash}p{1.0cm}}
\hline
\textbf{SOC level} & \textbf{$a_1$} & \textbf{$a_2$} & \textbf{$a_3$} & rmse [m$\Omega$] \\
\hline
90\% SOC & 0.0023 & -0.0857 & 0.0001 & 0.0091 \\
65\% SOC & 0.0024 & -0.0866 & 0.0001 & 0.0092 \\
40\% SOC & 0.0024 & -0.0909 & 0.0001 & 0.0090 \\
\hline
\end{tabular}}
\label{table:resistance_temp_fit}
\end{table}

The dependence of resistance on temperature is further examined in Fig. \ref{fig:paper_reistances_temp}. For each SOC, three distinct clusters are observed, corresponding to the tested environmental temperatures (yellow: 15$^\circ$C, red: 25$^\circ$C, and brown: 35$^\circ$C). Each cluster includes all \textit{Cells} across the modules. The measured module temperature does not exactly match the setpoint due to the transient thermal dynamics during the test, with deviations of up to $+4^\circ$C (see Appendix III). To account for this, each data point is associated with the average module temperature measured over the eight current pulses at a given SOC. The clusters exhibit  different resistance levels, with higher temperatures corresponding to lower resistance, consistent with prior findings \cite{barcellona2022aging}. The fitted coefficients of the resistance–temperature relationship in \eqref{eq:r_T_relation} are reported in Table \ref{table:resistance_temp_fit}. The resulting root mean square error (rmse) is on the order of 0.009 m$\Omega$, i.e., two orders of magnitude smaller than the absolute resistance values, indicating high fitting accuracy. These results highlight the strong influence of temperature on cell electrical behavior and provide a plausible explanation for the observed resistance heterogeneity within modules. In particular, the results suggest the presence of intra-module thermal gradients, where \textit{Cell} 2 may operate at a higher temperature than \textit{Cells} 1 and 3, leading to lower resistance. Such gradients may arise from the module layout, non-uniform heat generation, and thermal coupling among \textit{Cells}. We note, however, that module-level (surface) temperature measurements alone are insufficient to infer (or exclude) \textit{Cell}-level thermal gradients, as the casing may smooth out localized internal temperature variations (see Appendix III).
Finally, we note that, due to the adopted voltage measurement configuration, the computed internal resistance inherently includes voltage drops across module interconnect resistances. As these contributions cannot be separated from the \textit{Cells}’ terminal voltages with the present setup, their exact impact on the reported resistance values and associated heterogeneity cannot be quantified; however, it is expected to be small.
If interconnect resistances were uniform within a module, they by themselves would not explain the observed resistance heterogeneity, suggesting that geometry-dependent interconnect effects are unlikely to be the dominant factor. Nevertheless, this cannot be conclusively verified with the current measurements. Further investigation, particularly involving \textit{Cell}-level temperature sensing and dedicated interconnect resistance characterization, is therefore required to fully explain this behavior.

\section{CONCLUSION} \label{Conclusion}
The experimental platform and methodology presented in this paper provide a non-invasive approach for cell screening during module testing. We designed and built a dedicated low-power, Monitoring \& Balancing Hardware (M\&BH) platform with two primary functionalities:  (i) measuring and digitizing the voltages of the internal \textit{Cells}, specifically, three voltage readings corresponding to the three series-connected \textit{Cells}, and communicating these measurements to the battery cycler for data logging; and (ii) performing cell balancing during cycling. The balancing circuitry is critical for maintaining cell integrity during testing, preventing voltage imbalances that could compromise cell operation during module-level experiments. The collected signals are analyzed to extract quantitative metrics for assessing \textit{Cell} performance. Specifically, capacity tests are used to determine the discharge capacity and energy of each \textit{Cell}, and to evaluate how these quantities propagate to the module-level. In addition, HPPC tests are used to compute \textit{Cell} internal resistance, which is characterized across different SOC values and module temperatures. The extracted metrics provide a diagnostic view of the pack’s internal components, offering insights into state of health and intra-pack heterogeneity. The proposed methodology identifies distinct patterns within each module: the first \textit{Cell} in the series is most often the weakest in terms of capacity, while the intermediate \textit{Cell} consistently exhibits the lowest resistance. Moreover, the proposed module sensing setup enables studying trends between diagnostic indicators and operating variables such as temperature, allowing the identification and fitting of analytical relationships (e.g., resistance versus temperature). Future work will focus on direct \textit{Cell} temperature measurements, which are expected to provide a clearer understanding of potential intra-module thermal gradients and their impact on resistance heterogeneity across \textit{Cells}. Broadly, this work is particularly relevant for second-life applications, where accurately assessing the internal state of battery modules is essential for safe reuse and repurposing, especially in the presence of cell-to-cell heterogeneities accumulated during first-life operation. In this context, the proposed M\&BH platform can serve as an accurate, non-disruptive, and safe screening tool to support module qualification and sorting prior to second-life deployment.

%\section*{DECLARATION OF COMPETING INTERESTS}
%William A. Paxton, employed by VW Group of America, Inc., provided the battery modules used in the experiments described in this paper to collect the data for this study. This relationship could be perceived as a potential conflict of interest.

\section*{ACKNOWLEDGMENT}
The authors would like to thank VW Group of America, Inc., for providing the battery pack used as the case study in this work and for performing the pack disassembly operation through which the tested modules were extracted. The authors also thank P. Bosoni (pbosoni@stanford.edu) for his help with the experimental module testing activity.

\section*{APPENDIX I: MEASUREMENT AND DATA ACQUISITION}
With reference to the M\&BH schematics in Fig.~\ref{fig:photo_setup}(d), we further detail the voltage measurement pipeline.

\subsection{VOLTAGE SCALING AND FILTERING}
\begin{figure}[t]
    \centering
    \includegraphics[width=1\linewidth]{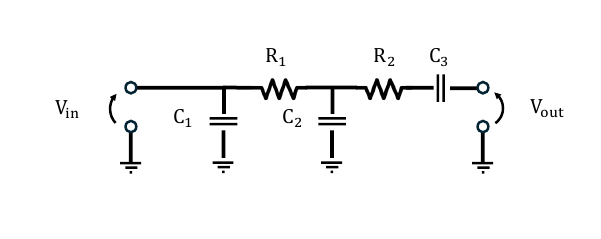}
    \vspace{-4.0em}
    \caption{\textbf{LP Filter scheme.} The circuit comprises resistors $R_1$ and $R_2$, and capacitors $C_1$, $C_2$ and $C_3$, with parameter values listed in Table \ref{table:board_specs}.}
    \label{fig:analog_filter}
\end{figure}

The node potentials C1+, C2+, and C3+ are directly accessed through the module external connectors, defined as follows:
\begin{itemize}
\item C1+: potential at the positive terminal of \textit{Cell} 1 with respect to the common ground G-;
\item C2+: potential at the positive terminal of \textit{Cell} 2 with respect to the common ground G-;
\item C3+: potential at the positive terminal of \textit{Cell} 3 with respect to the common ground G-.
\end{itemize}
Under this measurement configuration, the acquired voltages inherently include contributions from interconnect resistances between \textit{Cells}, which cannot be decoupled from the individual \textit{Cell} terminal voltages. However, these contributions are expected to be negligible relative to the \textit{Cell} voltage.
Each node potential is scaled by a factor of 4:1 using voltage dividers formed by $R_H$ and $R_L$ (with scaling factor $\frac{R_H}{R_H+R_L}$), ensuring compatibility with the ADC input range.
To attenuate high-frequency noise and mitigate aliasing effects, a filtering network is placed between each divider output and the corresponding ADC channel, resulting in three identical low-pass filters.
The filter topology is shown in Fig.~\ref{fig:analog_filter}, where $V_{in}$ denotes the voltage divider output and $V_{out}$ the voltage applied to the individual ADC input channel.

The filter comprises three functional blocks:
\begin{itemize}
    \item input shunt capacitor ($C_1$): provides a local high-frequency return path to ground, reducing fast transients at the filter input;
    \item low-pass stage ($R_1, C_2$): defines the measurement bandwidth $f_{LP}$ of the filter;
    \item DC blocking ($R_2, C_3$): acts as an open-circuit to direct-current (DC), preventing continuous current flow into the ADC to protect the input terminal.
\end{itemize}

\begin{figure*}[h]
\centering
\includegraphics[width=0.85\linewidth]{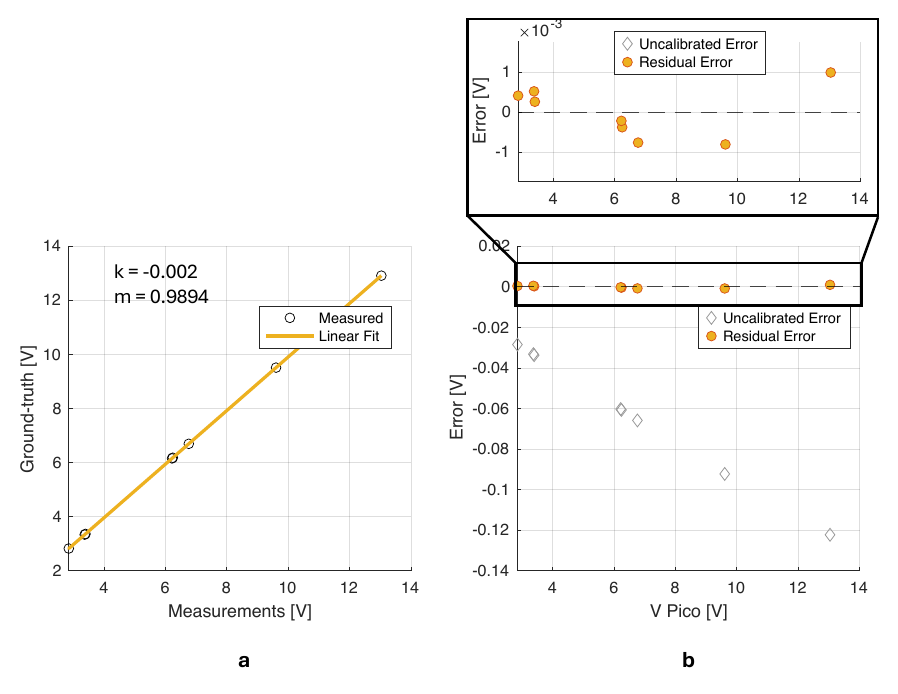}
\vspace{-1.5em}
\caption{\textbf{ADC channel calibration.} Calibration results for channel 1. (\textbf{a}) Measured voltage from the circuit versus oscilloscope ground truth, with the best linear fit obtained via ordinary least squares.  The fitted gain and bias indicate that only minor correction is required. (\textbf{b}) Measurement errors before and after calibration, comparing the uncalibrated error (ADC measurement vs. ground truth) and the calibration residual error.  While the uncalibrated error exhibits a diverging trend, the calibrated error remains within 1 mV.}
\label{fig:Calibration}
\end{figure*}

In the following, the cutoff frequency of the filtering network is computed by modeling $V_{in}$ as an ideal voltage source and  $V_{out}$ as an open (high-impedance) circuit.
Under these assumptions, the input shunt capacitor is directly driven by the input voltage and does not constitute an independent state, while the DC-blocking capacitor does not affect the output dynamics due to the absence of a current path in that branch.
Under these assumptions, the input--output transfer function reduces to:
\begin{equation}
    H(s)=\frac{1}{R_1 C_2s+1},
\end{equation}
with unitary gain $H(0)=1$, time constant $\tau=R_1 C_2$, and cutoff frequency $f_{LP}=\frac{1}{2 \pi \tau}$ (in Hz).
Using the component values listed in Table \ref{table:board_specs}, the resulting cutoff frequency is $f_{LP}=112.9$ Hz (i.e., $\tau=0.00141$ s).
Relaxing the ideal assumptions introduces higher-order dynamics:
\begin{itemize}
    \item a non-ideal voltage source (with a series resistance) causes the input shunt capacitor to introduce an additional pole;
    \item a finite output load enables causes $C_3$ to introduce a zero in the transfer function, leading the overall network toward a band-pass behavior.
\end{itemize}
The selected component values yield an overall filtering response consistent with standard on-board filtering stages.

\begin{table}[!h]
\centering
\caption{Analog electrical circuit component values.}
\resizebox{0.43\columnwidth}{!}{%
\begin{tabular}{lc}
\hline
\textbf{Component} & \textbf{Value} \\ \hline
Resistor $R_b$ & $67.5~\Omega$ \\
Resistor $R_H$ & $30~\text{k}\Omega$ \\
Resistor $R_L$ & $10~\text{k}\Omega$ \\
Resistor $R_1$ & $3~\text{k}\Omega$ \\
Resistor $R_2$ & $2~\text{k}\Omega$ \\
Capacitor $C_1$ & $47~\text{nF}$ \\
Capacitor $C_2$ & $470~\text{nF}$ \\
Capacitor $C_3$ & $47~\text{nF}$ \\ \hline
\end{tabular}}
\label{table:board_specs}
\end{table}

\subsection{ADC, DIGITAL CORRECTION, AND LATENCY}
The voltage measurement circuit utilizes a multi-channel ADS1115 as ADC (allowing up to four input channels), to sample the three distinct voltage channels at a frequency of $10\text{ Hz}$ (sampling period $T_s = 0.1\text{ s}$).
Each voltage channel represents the scaled and low-passed version of the initial node potentials (C1+, C2+ and C3+).
The ADC is configured with a 16-bit resolution and a programmable gain amplifier setting of $\pm 4.096\,\text{V}$, which sets the maximum allowable input voltage range.
In our implementation, we conservatively selected a 4:1 voltage divider ratio to remain within this limit with margin; with a maximum module voltage of $12.6\,\text{V}$, the ADC input reaches at most $\approx 3.15\,\text{V}$.
For single-ended measurements, the ADS1115 provides 15 bits of usable resolution ($2^{15} = 32,768$ levels).
The native quantization step for each ADC channel is therefore calculated as:
\[ V_{\text{quantization,native}} = \frac{4.096\text{ V}}{2^{15}} = 125\text{ \textmu V}. \]
The ADC communicates with the micro-controller (Raspberry Pi Pico) through a I2C line.
The micro-controller performs four fundamental tasks related to voltage measurement.

\begin{figure}[!b]
    \centering
    \includegraphics[width=1.0\linewidth]{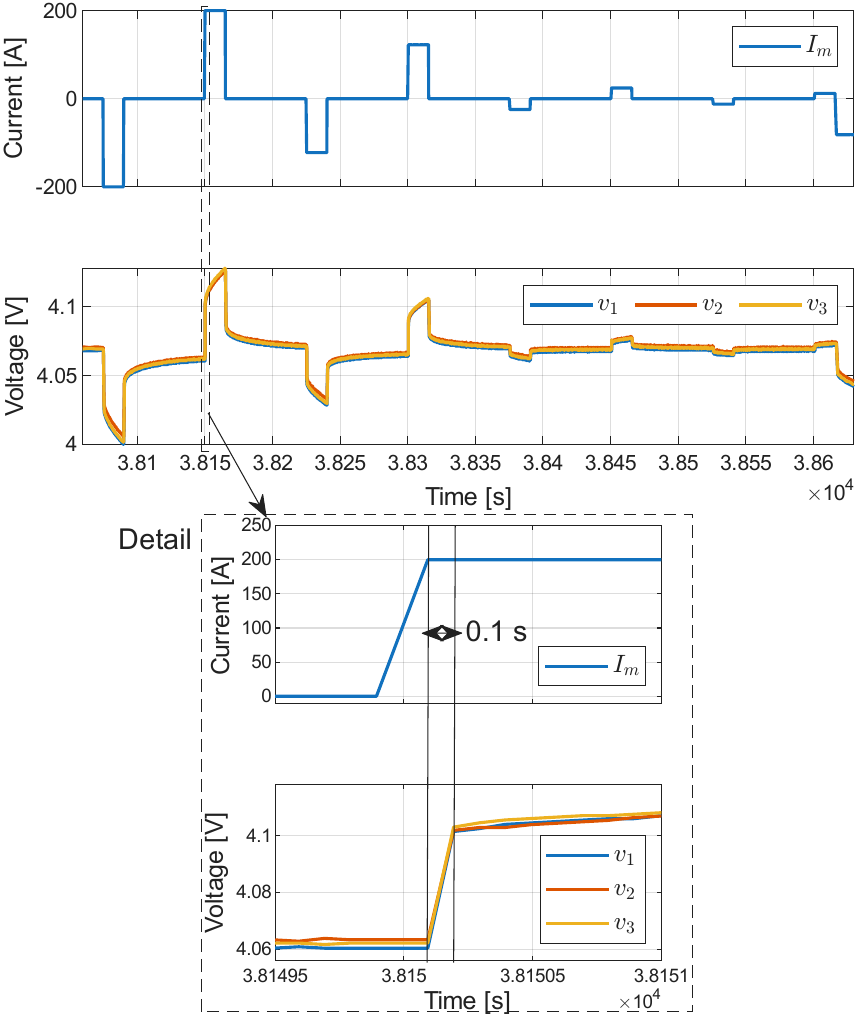}
    %\vspace{-1.0em}
    \caption{\textbf{Latency visualization.} The top two plots report the module current and \textit{Cell} voltages during the HPPC test at 90\% SOC, while the bottom two plots show a zoomed-in view of the CAN communication latency. The instantaneous voltage response measured by the cycler is observed 100 ms after the application of the current pulse.}
    \label{fig:Latency}
\end{figure}

\subsubsection{RESCALING} 
It rescales the potentials sampled by the ADC to their original voltage range, i.e., multiplying them by 4.
In this sense, the effective voltage quantization error for each ADC channel becomes:
\[ V_{\text{quantization,effective}} = 125\text{ \textmu V} \times 4 = 0.5\text{ mV} . \]

\subsubsection{SENSING CORRECTION} 
To compensate for hardware tolerances in the voltage dividers and the ADC internal gain, the data from the $j$-th ADC channel (already multiplied by 4) undergo a linear correction:
\[ V_\text{corrected}^{\text{channel }j}=m_j \cdot V_\text{not corrected}^{\text{channel }j} +k_j , \]
where $m_j$ is the correction gain and $k_j$ is the correction bias.
We consider a correction model for each ADC channel.
The parameter pair $(m_j,k_j)$ for each channel is identified through a calibration procedure in which the ground-truth corrected voltage is provided by a high-precision digital oscilloscope.
The calibration dataset is obtained by measuring the terminal voltage of randomly selected cells from our laboratory and connecting them in series to span input voltages up to 12 V.
Fig. \ref{fig:Calibration} shows that the linear correction model (best linear fitting using ordinary least square) yields model residuals smaller than $1$~mV across the whole voltage operating range.
The model fittings for the three ADC channels are reported in Table \ref{tab:correction_model_specs}.

\begin{table}[!h]
\centering
\caption{Fitted linear correction models for the ADC channels.}
\resizebox{0.75\columnwidth}{!}{%
\begin{tabular}{lcc}
\hline
\textbf{ADC channel} & \textbf{gain $m$ [$-$]} & \textbf{bias $k$ [V]} \\ \hline
First channel & $0.9894$  & $-0.002$  \\
Second channel & $0.9910$ & $-0.002$  \\
Third channel & $0.9907$ & $-0.002$ \\ \hline
\end{tabular}}
\label{tab:correction_model_specs}
\end{table}

\subsubsection{\textit{Cell} VOLTAGE COMPUTATION} 
The \textit{Cell} voltages are computed starting from the corrected channel potentials:
\begin{itemize}
        \item \textit{Cell} 1 voltage: $v_1=V_\text{corrected}^\text{channel 1}$;
        \item \textit{Cell} 2 voltage: $v_2=V_\text{corrected}^\text{channel 2}-V_\text{corrected}^\text{channel 1}$;
        \item \textit{Cell} 3 voltage: $v_3=V_\text{corrected}^\text{channel 3}-V_\text{corrected}^\text{channel 2}$. 
\end{itemize}

\subsubsection{COMMUNICATION WITH CYCLER} 
The \textit{Cell} voltage values are transmitted to the cycler via the CAN bus at a sampling frequency of 10 Hz (consistent with the cycler maximum logging frequency).
These values are used by the cycler (i) for signal logging and (ii) to enforce current cutoffs based on prescribed \textit{Cell} voltage thresholds during the tests (see Fig. \ref{fig:current_profile_1} for details).
Communication is bidirectional: over the same CAN link, the Arbin sends single-bit commands to the micro-controller to disable cell balancing (i.e., during discharge experiments) or enable it (i.e., to rebalance the \textit{Cells} during CC-CV charging).
Due to CAN bus communication, the voltage measurements are recorded by the battery cycler with an inherent latency observed to be approximately 100 ms.
This delay, which corresponds to one sampling period $T_s$, is illustrated in Fig.~\ref{fig:Latency} for an HPPC current pulse.
As shown, the instantaneous voltage drop induced by the current step is registered by the cycler about 100 ms after the step change in current.

\section*{APPENDIX II: \textit{Cell} DYNAMICS UNDER BALANCING}
The electrical behavior of $j$-th \textit{Cell} of the module is modeled using a zero-order equivalent circuit model (ECM): 
\begin{equation}
    v_j = OCV_j(SOC_j) + R_{0,j} i_j,\label{eq:ECM0th}
\end{equation}
with $OCV_j$ open-circuit voltage, $R_{0,j}$ ohmic resistance and $i_j$ the current flowing through the \textit{Cell}. In this work, we follow the Arbin sign convention, where the current is positive during charging; negative when discharging. The open-circuit voltage is a function of the \textit{Cell}'s state-of-charge $SOC_j$, given by Coulomb counting:
\begin{equation}
    \dot{SOC}_j = \frac{i_j}{Q_j},\label{eq:SOC_coulomb}
\end{equation}
with $Q_j$ the \textit{Cell}'s nominal capacity. According to the circuit schematic in Fig. \ref{fig:photo_setup}(d), in parallel with the $j$-th \textit{Cell} is a branch containing the switch $S_j$ and the bleeding resistor $R_b$, whose value is the same for all \textit{Cells}. Notice that the switch position of each parallel branch is controlled by the balancing logic in \eqref{eq:balancing_alg}. With reference to the $j$-th \textit{Cell}, two scenarios arise.

\setcounter{subsubsection}{0}
\subsubsection{THE SWITCH IS OPEN}
In this scenario, the parallel branch is an open-circuit and no current flows through the parallel branch. Therefore, the current flowing through the \textit{Cell} is the same as the module current: 
\begin{equation}
    \text{(open switch)} \ i_j=I_m, \ \forall j\in\{1,2,3\}.\label{eq:ij=Im}
\end{equation}
By plugging \eqref{eq:ij=Im} into \eqref{eq:SOC_coulomb}, all \textit{Cells} undergo the same SOC evolution, up to differences arising only from their individual capacities.

\subsubsection{THE SWITCH IS CLOSED}
In this scenario, the module current is partially diverted into the parallel branch, therefore less current flows through the \textit{Cell}. Considering that the voltage divider resistances $R_H$ and $R_L$ are much greater than the bleeding resistance $R_b$ (consistently with the values listed in Table \ref{table:board_specs}), the following relations for the $j$-th parallel branch hold (from the Kirchhoff's laws):
\begin{equation}
    \text{(closed switch)} \ \begin{cases}
        I_m=i_j+i_{R_{b}} \\
        v_j=v_{R_{b}} \\
        v_{R_{b}}=R_{b}i_{R_{b}}
    \end{cases}.\label{eq:parall_branch}
\end{equation}
Here, $v_{R_{b}}$ is the voltage drop across the bleeding resistor, and $i_{R_{b}}$ the current flowing through the parallel branch. By incorporating \eqref{eq:ECM0th}-\eqref{eq:SOC_coulomb},\eqref{eq:parall_branch}, we obtain the following SOC dynamic equation for the $j$-th \textit{Cell}:
\begin{equation}
    \dot{SOC}_j=\frac{R_{b}}{Q_j(R_{0,j}+R_{b})}I_m-\frac{OCV_j(SOC_j)}{Q_j(R_{0,j}+R_{b})}.
    \label{eq:SOC_bal}
\end{equation}
Therefore, each \textit{Cell}'s SOC can be controlled independently via the closing of the corresponding parallel switch. Two key observations are made on the value of $R_b$:
\begin{itemize}
    \item given a constant module current, if the bleeding resistance value is exactly equal to $\frac{OCV_j(SOC_j)}{I_m}$, then no current flows through the $j$-th \textit{Cell}, effectively freezing its state-of-charge, i.e., $\dot{SOC}_j=0$;
    \item in the case of bleeding resistance with value $R_b \gg R_{0,j}$, only a very small fraction of the module current is diverted into the parallel branch. Consequently, the evolution of the \textit{Cell}'s SOC is only marginally slowed down with balancing. 
\end{itemize}
Generally, $R_b$ may be selected based on the desired SOC dynamics under balancing. In this context, $R_b$ serves as a design parameter: the lower its value, the smaller the \textit{Cell} charging rate.

\section*{APPENDIX III: TEMPERATURE INSTRUMENTATION}
\begin{figure}[!t]
    \centering
    \includegraphics[width=1.0\linewidth]{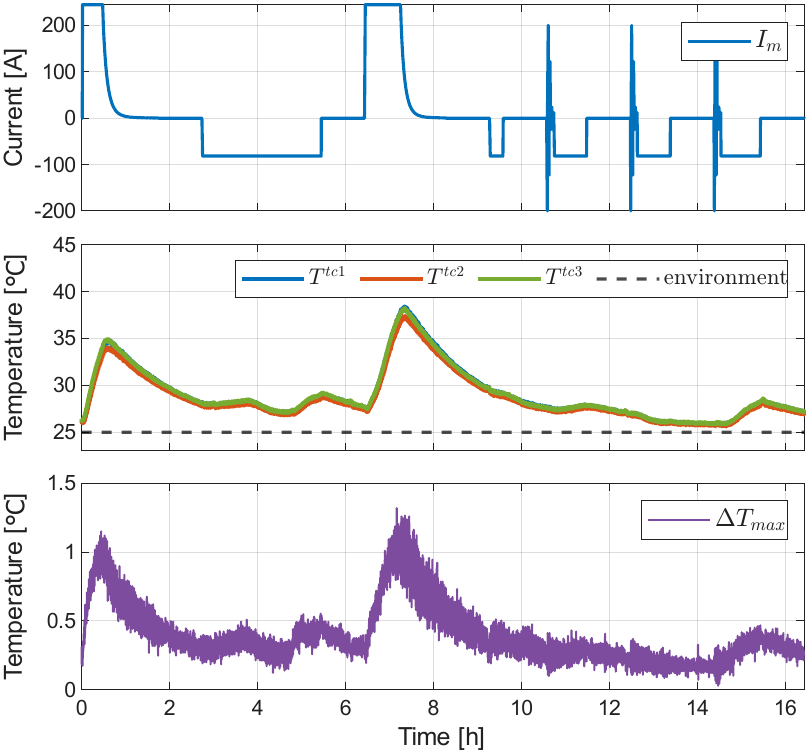}
    %\vspace{-1.0em}
    \caption{Top: module current over the entire test. Middle: thermocouple temperatures $T^{tc1}$, $T^{tc2}$, $T^{tc3}$ (solid lines) with environmental setpoint (dashed). Bottom: maximum surface temperature difference $\Delta T_{max}$.}  \label{fig:thermocouples}
\end{figure}

Fig. \ref{fig:photo_setup}(b) shows the module instrumented with three Type-T thermocouples (model 5SRTC-TT-T) placed near the positive terminal, negative terminal, and center of the upper surface.
The corresponding temperature traces ($T^{tc1}$, $T^{tc2}$, $T^{tc3}$) for module no.29 are reported in Fig. \ref{fig:thermocouples} at a chamber setpoint of $25^\circ$C.
The module starts at thermal equilibrium, and its surface temperature increases with current, peaking during 1C charging. During rest periods, the temperature relaxes toward the setpoint but does not fully return to equilibrium due to limited rest duration.
The three thermocouples report consistent readings throughout the test.
Defining the maximum module surface temperature difference as:
\begin{equation}
    \Delta T_{max}=\max(T^{tc1},T^{tc2},T^{tc3})-\min(T^{tc1},T^{tc2},T^{tc3}),
\end{equation}
the bottom subplot of Fig. \ref{fig:thermocouples} shows that $\Delta T_{max}$ reaches $\sim1^\circ$C during 1C charge and remains below $0.5^\circ$C during the C/3 test and HPPC pulses.
This indicates weak thermal gradients across the module surface\footnote{Differences below $1^\circ$C are within the $\pm 1^\circ$C thermocouple accuracy \cite{TypeTArbin}.}. 
Despite the observed \textit{Cell} internal resistance heterogeneity, the small $\Delta T_{max}$ alone does not explain the pattern in Fig. \ref{fig:paper_reistances}.
Surface measurements may not capture internal \textit{Cell} temperatures, as the casing can smooth localized gradients.
Therefore, intra-module thermal gradients at the \textit{Cell}-level cannot be excluded based on surface data alone.
Further investigation should rely on direct \textit{Cell}-level temperature measurements.

\bibliographystyle{Bibliography/IEEEtranTIE}
\bibliography{Bibliography/BIB_xx-TIE-xxxx}\ %IEEEabrv instead of IEEEfull

@article{catenaro2021experimental,
  title={Experimental data of lithium-ion batteries under galvanostatic discharge tests at different rates and temperatures of operation},
  author={Catenaro, Edoardo and Onori, Simona},
  journal={Data in Brief},
  volume={35},
  pages={106894},
  year={2021},
  publisher={Elsevier}
}

@article{moy2024second,
  title={Second-life lithium-ion battery aging dataset based on grid storage cycling},
  author={Moy, Kevin and Khan, Muhammad Aadil and Fasolato, Simone and Pozzato, Gabriele and Allam, Anirudh and Onori, Simona},
  journal={Data in Brief},
  volume={57},
  pages={111046},
  year={2024},
  publisher={Elsevier}
}

@article{cui2024taking,
  title={Taking second-life batteries from exhausted to empowered using experiments, data analysis, and health estimation},
  author={Cui, Xiaofan and Khan, Muhammad Aadil and Pozzato, Gabriele and Singh, Surinder and Sharma, Ratnesh and Onori, Simona},
  journal={Cell Reports Physical Science},
  volume={5},
  number={5},
  year={2024},
  publisher={Elsevier}
}

@article{preger2025impact,
  title={Impact of module configuration on lithium-ion battery performance and degradation: Part i. energy throughput, voltage spread, and current distribution},
  author={Preger, Yuliya and Mueller, Jacob and Fresquez, Armando and Allu, Srikanth and Rich, Chaz},
  journal={Journal of The Electrochemical Society},
  volume={172},
  number={5},
  pages={050540},
  year={2025},
  publisher={IOP Publishing}
}

@inproceedings{niu2024research,
  title={Research on Digital Calibration of Voltage Measurement for Automotive Battery Management Systems},
  author={Niu, Han and Luo, Bingyin},
  booktitle={2024 13th International Conference on Communications, Circuits and Systems (ICCCAS)},
  pages={122--126},
  year={2024},
  organization={IEEE}
}

@article{kurkin2025battery,
  title={Battery management system for electric vehicles: comprehensive review of circuitry configuration and algorithms},
  author={Kurkin, Andrey and Chivenkov, Alexander and Aleshin, Dmitriy and Trofimov, Ivan and Shalukho, Andrey and Vilkov, Danil},
  journal={World Electric Vehicle Journal},
  volume={16},
  number={8},
  pages={451},
  year={2025},
  publisher={MDPI}
}

@article{azimi2022extending,
  title={Extending life of lithium-ion battery systems by embracing heterogeneities via an optimal control-based active balancing strategy},
  author={Azimi, Vahid and Allam, Anirudh and Onori, Simona},
  journal={IEEE Transactions on Control Systems Technology},
  volume={31},
  number={3},
  pages={1235--1249},
  year={2022},
  publisher={IEEE}
}

@article{gao2024evaluation,
  title={Evaluation of the second-life potential of the first-generation Nissan Leaf battery packs in energy storage systems},
  author={Gao, Wei and Cao, Zhi and Kurdkandi, Naser Vosoughi and Fu, Yuhong and Mi, Chirs},
  journal={ETransportation},
  volume={20},
  pages={100313},
  year={2024},
  publisher={Elsevier}
}

@article{bach2024fair,
  title={Fair market valuation of electric vehicle batteries in second life applications},
  author={Bach, Amadeus and Reichelstein, Stefan and Onori, Simona and Zhuang, Jihan},
  journal={Available at SSRN 4940320},
  year={2024}
}

@article{zhuang2025technoeconomic,
  title={Technoeconomic decision support for second-life batteries},
  author={Zhuang, Jihan and Bach, Amadeus and van Vlijmen, Bruis HC and Reichelstein, Stefan J and Chueh, William and Onori, Simona and Benson, Sally M},
  journal={Applied Energy},
  volume={390},
  pages={125800},
  year={2025},
  publisher={Elsevier}
}

@article{muratori2025trends,
  title={Trends and 2025 insights on the rise of electric vehicles in the USA},
  author={Muratori, Matteo and Arent, Doug and Bazilian, Morgan D and Bistline, John and Borlaug, Brennan and Brown, Austin and Cazzola, Pierpaolo and Dede, Ercan M and Gearhart, Chris and Greene, David and others},
  journal={Nature Reviews Clean Technology},
  pages={1--19},
  year={2025},
  publisher={Nature Publishing Group UK London}
}

@article{barcellona2022aging,
  title={Aging effect on the variation of Li-ion battery resistance as function of temperature and state of charge},
  author={Barcellona, Simone and Colnago, Silvia and Dotelli, Giovanni and Latorrata, Saverio and Piegari, Luigi},
  journal={Journal of Energy Storage},
  volume={50},
  pages={104658},
  year={2022},
  publisher={Elsevier}
}

@misc{TypeTArbin,
  title			= "Arbin Instruments, Auxiliary Temperature Measurement", 
  note			= "\url{https://www.arbin.com/auxiliary-temperature-measurement.html}"
}

@inproceedings{lee2011comparison,
  title={Comparison of passive cell balancing and active cell balancing for automotive batteries},
  author={Lee, Wai Chung and Drury, David and Mellor, Phil},
  booktitle={2011 IEEE Vehicle Power and Propulsion Conference},
  pages={1--7},
  year={2011},
  organization={IEEE}
}

@article{weng2024current,
  title={Current imbalance in dissimilar parallel-connected batteries and the fate of degradation convergence},
  author={Weng, Andrew and Movahedi, Hamidreza and Wong, Clement and Siegel, Jason B and Stefanopoulou, Anna},
  journal={Journal of Dynamic Systems, Measurement, and Control},
  volume={146},
  number={1},
  pages={011106},
  year={2024},
  publisher={American Society of Mechanical Engineers}
}

@article{partcap,
author={Mohamed Ahmeid and Musbahu Muhammad and Simon Lambert and Pierrot S. Attidekou and Zoran Milojevic},
journal={Journal of Energy Storage},
title={A rapid capacity evaluation of retired electric vehicle battery modules using partial discharge test},
year={2022},
month= {June},
volume={50},
}

@article{intro,
title = {Fast screening of lithium-ion batteries for second use with pack-level testing and machine learning},
journal = {eTransportation},
volume = {17},
pages = {100255},
year = {2023},
month= {July},
issn = {2590-1168},
author = {Sijia Yang and Caiping Zhang and Jiuchun Jiang and Weige Zhang and Haoze Chen and Yan Jiang and Dirk Uwe Sauer and Weihan Li},
}

@article{thermal,
title={Experimental analysis of thermal runaway and propagation in lithium-ion battery modules},
author={Lopez, Carlos F and Jeevarajan, Judith A and Mukherjee, Partha P},
journal={Journal of the electrochemical society},
volume={162},
number={9},
pages={A1905},
year={2015},
month={July},
publisher={IOP Publishing}
}

@misc{bibra2022global,
title		= {Global EV outlook 2022: Securing supplies for an electric future},
author		= {Bibra, Ekta Meena and Connelly, Elizabeth and Dhir, Shobhan and Drtil, Michael and Henriot, Pauline and Hwang, Inchan and Le Marois, Jean-Baptiste and McBain, Sarah and Paoli, Leonardo and Teter, Jacob},
year		= {2022},
url 		= {https://www.iea.org/reports/global-ev-outlook-2022}
}

@inbook{LiIonApp,
author 		= {Deng, Haokun and Aifantis, Katerina E.},
publisher 	= {John Wiley and Sons, Ltd},
isbn 		= {9783527836703},
title 		= {Applications of Lithium Batteries},
booktitle 	= {Rechargeable Ion Batteries},
chapter 	= {4},
pages 		= {83-103},
year 		= {2023},
}

@misc{mod_test,
title		= {Mastering Battery Testing: Everything you Need to Know},
author		= {Averna},
year		= {2024},
url 		= {https://www.averna.com/mastering-battery-testing-everything-you-need-to-know}
}

@book{pesaran1997thermal,
title={Thermal performance of EV and HEV battery modules and packs},
author={Pesaran, Ahmad A and Vlahinos, Andreas and Burch, Steven D and others},
year={1997},
publisher={National Renewable Energy Laboratory Golden, CO, USA}
}

@article{PIOMBO2024110783,
title = {Unveiling the performance impact of module level features on parallel-connected lithium-ion cells via explainable machine learning techniques on a full factorial design of experiments},
journal = {Journal of Energy Storage},
volume = {84},
pages = {110783},
year = {2024},
issn = {2352-152X},
author = {Gabriele Piombo and Simone Fasolato and Robert Heymer and Marc Hidalgo and Mona {Faraji Niri} and Simona Onori and James Marco},
}

@article{PIOMBO2024110227,
title = {Full factorial design of experiments dataset for parallel-connected lithium-ion cells imbalanced performance investigation},
journal = {Data in Brief},
volume = {53},
pages = {110227},
year = {2024},
issn = {2352-3409},
author = {Gabriele Piombo and Simone Fasolato and Robert Heymer and Marc F. Hidalgo and Mona {Faraji Niri} and Davide M. Raimondo and James Marco and Simona Onori},
}

@article{BAUMANN2018295,
title = {Parameter variations within Li-Ion battery packs: Theoretical investigations and experimental quantification},
journal = {Journal of Energy Storage},
volume = {18},
pages = {295-307},
year = {2018},
issn = {2352-152X},
author = {Michael Baumann and Leo Wildfeuer and Stephan Rohr and Markus Lienkamp},
}

@article{WASSILIADIS2022100167,
title = {Quantifying the state of the art of electric powertrains in battery electric vehicles: Range, efficiency, and lifetime from component to system level of the Volkswagen ID.3},
journal = {eTransportation},
volume = {12},
pages = {100167},
year = {2022},
issn = {2590-1168},
author = {Nikolaos Wassiliadis and Matthias Steinstr{\"a}ter and Markus Schreiber and Philipp Rosner and Lorenzo Nicoletti and Florian Schmid and Manuel Ank and Olaf Teichert and Leo Wildfeuer and Jakob Schneider and Alexander Koch and Adrian K{\"o}nig and Andreas Glatz and Josef Gandlgruber and Thomas Kr{\"o}ger and Xue Lin and Markus Lienkamp},
}

@Article{TEMP1Fill,
AUTHOR = {Fill, Alexander and Mader, Tobias and Schmidt, Tobias and Llorente, Raphael and Birke, Kai Peter},
TITLE = {Measuring Test Bench with Adjustable Thermal Connection of Cells to Their Neighbors and a New Model Approach for Parallel-Connected Cells},
JOURNAL = {Batteries},
VOLUME = {6},
YEAR = {2020},
NUMBER = {1},
ARTICLE-NUMBER = {2},
ISSN = {2313-0105}
}

@article{TEMP2JOCHER,
title = {A novel measurement technique for parallel-connected lithium-ion cells with controllable interconnection resistance},
journal = {Journal of Power Sources},
volume = {503},
pages = {230030},
year = {2021},
issn = {0378-7753},
author = {P. Jocher and M. Steinhardt and S. Ludwig and M. Schindler and J. Martin and A. Jossen},
}

@article{TEMP3LI,
title = {Effect of parallel connection topology on air-cooled lithium-ion battery module: Inconsistency analysis and comprehensive evaluation},
journal = {Applied Energy},
volume = {313},
pages = {118758},
year = {2022},
issn = {0306-2619},
author = {Changlong Li and Naxin Cui and Long Chang and Zhongrui Cui and Haitao Yuan and Chenghui Zhang}
}

@article{C2C1DIAO,
title = {Management of imbalances in parallel-connected lithium-ion battery packs},
journal = {Journal of Energy Storage},
volume = {24},
pages = {100781},
year = {2019},
issn = {2352-152X},
author = {Weiping Diao and Michael Pecht and Tao Liu},
}

@article{C2C2TIAN,
title = {Parallel-connected battery module modeling based on physical characteristics in multiple domains and heterogeneous characteristic analysis},
journal = {Energy},
volume = {239},
pages = {122181},
year = {2022},
issn = {0360-5442},
author = {Yong Tian and Zhijia Huang and Xiaoyu Li and Jindong Tian},
}

@article{TOP1GRUN,
title = {Influence of circuit design on load distribution and performance of parallel-connected Lithium ion cells for photovoltaic home storage systems},
journal = {Journal of Energy Storage},
volume = {17},
pages = {367-382},
year = {2018},
issn = {2352-152X},
author = {Thorsten Gr{\"u}n and Kevin Stella and Olaf Wollersheim},
}

@article{seriesZhou,
author={Zhou, Zhongkai and Duan, Bin and Kang, Yongzhe and Zhang, Qi and Shang, Yunlong and Zhang, Chenghui},
journal={IEEE Transactions on Transportation Electrification}, 
title={Online State of Health Estimation for Series-Connected {LiFePO}$_4$ Battery Pack Based on Differential Voltage and Inconsistency Analysis}, 
year={2024},
volume={10},
number={1},
pages={989-998},
}

@article{GU2024,
title = {Challenges and opportunities for second-life batteries: Key technologies and economy},
journal = {Renewable and Sustainable Energy Reviews},
volume = {192},
pages = {114191},
year = {2024},
issn = {1364-0321},
author = {Xubo Gu and Hanyu Bai and Xiaofan Cui and Juner Zhu and Weichao Zhuang and Zhaojian Li and Xiaosong Hu and Ziyou Song},
}

@inproceedings{allam2016characterization,
  title={Characterization of aging propagation in lithium-ion cells based on an electrochemical model},
  author={Allam, Anirudh and Onori, Simona},
  booktitle={2016 American Control Conference (ACC)},
  pages={3113--3118},
  year={2016},
  organization={IEEE}
}

@article{offer2012module,
  title={Module design and fault diagnosis in electric vehicle batteries},
  author={Offer, Gregory J and Yufit, Vladimir and Howey, David A and Wu, Billy and Brandon, Nigel P},
  journal={Journal of Power Sources},
  volume={206},
  pages={383--392},
  year={2012},
  publisher={Elsevier}
}

@inproceedings{wong2024differential,
  title={Differential Voltage Analysis and Patterns in Parallel-Connected Pairs of Imbalanced Cells},
  author={Wong, Clement and Weng, Andrew and Pannala, Sravan and Choi, Jeesoon and Siegel, Jason B and Stefanopoulou, Anna},
  booktitle={2024 American Control Conference (ACC)},
  pages={3492--3497},
  year={2024},
  organization={IEEE}
}

@article{naylor2024degradation,
  title={Degradation in parallel-connected lithium-ion battery packs under thermal gradients},
  author={Naylor Marlow, Max and Chen, Jingyi and Wu, Billy},
  journal={Communications Engineering},
  volume={3},
  number={1},
  pages={2},
  year={2024},
  publisher={Nature Publishing Group UK London}
}

\begin{IEEEbiography}[{\includegraphics[width=1in,height=1.25in,clip,keepaspectratio]{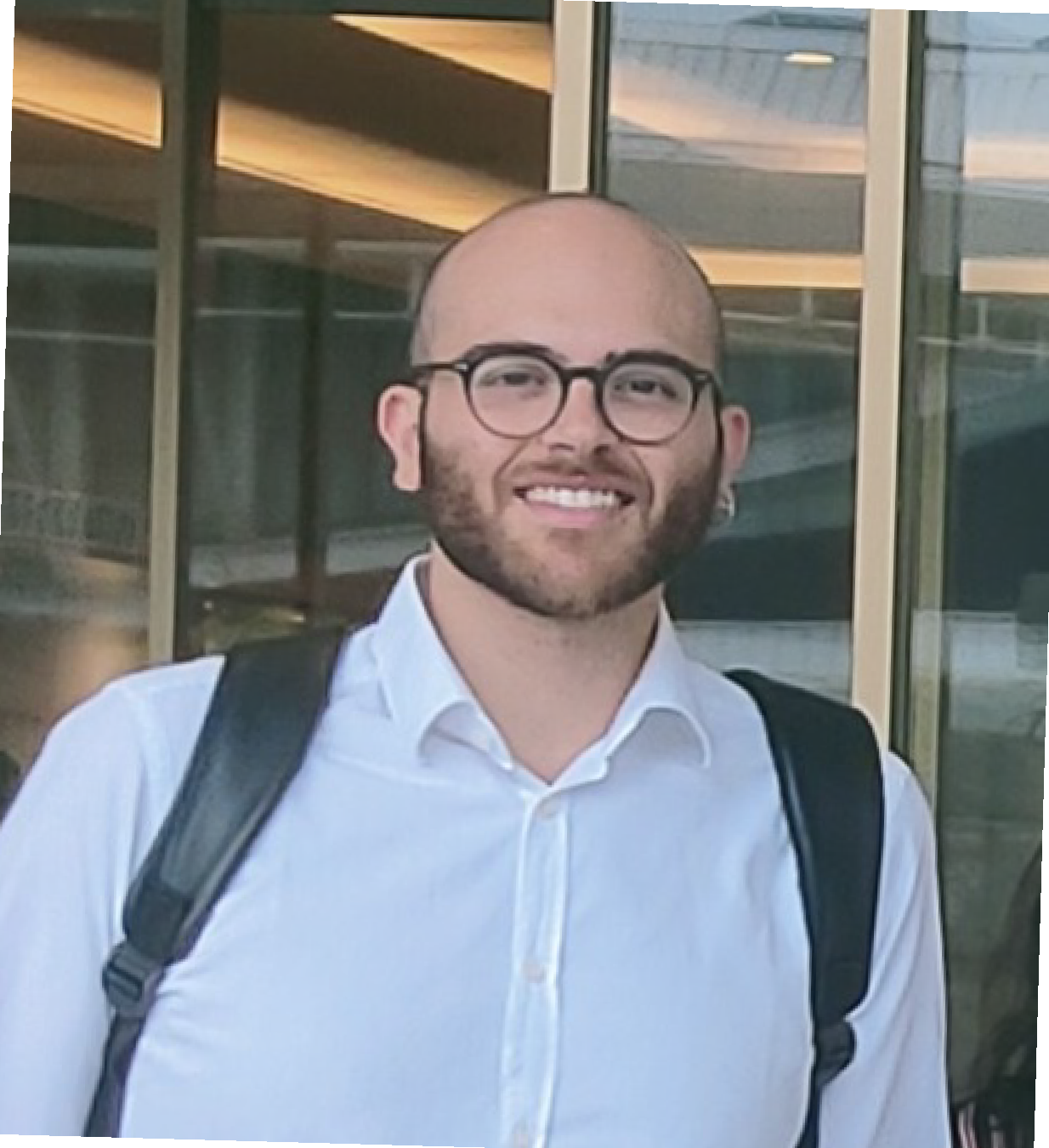}}]{GABRIELE MARINI } was born in Penne, Italy. He received the Ph.D. in Information Technology from Politecnico di Milano, Italy, in 2024. Since November 2024, he is a postdoctoral researcher at Stanford university, Department of Energy Science \& Engineering.

He has a background in systems, control and optimization theory. His current research focuses on state-of-health estimation of lithium-ion batteries, digital twins of EV battery packs, and parameter identification for battery electrochemical models. He has previously worked on control and estimation for transportation systems, particularly focusing on high-performance vehicles.
\end{IEEEbiography}

\begin{IEEEbiography}[{\includegraphics[width=1in,height=1.25in,clip,keepaspectratio]{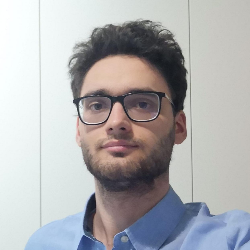}}]{ALESSANDRO COLOMBO } was born in Busto Arsizio, Italy. He received the B.Sc. and M.SC. in Automation and Control Engineering from Politecnico di Milano, Italy, in 2020 and 2022, respectively.

In December 2022, he started a PhD in Information Technology at Politecnico di Milano. From January to June 2024 he was a visiting student at Stanford university, at the Energy Science and Engineering department, where the work described in this paper took place.
\end{IEEEbiography}

\begin{IEEEbiography}[{\includegraphics[width=1in,height=1.25in,clip,keepaspectratio]{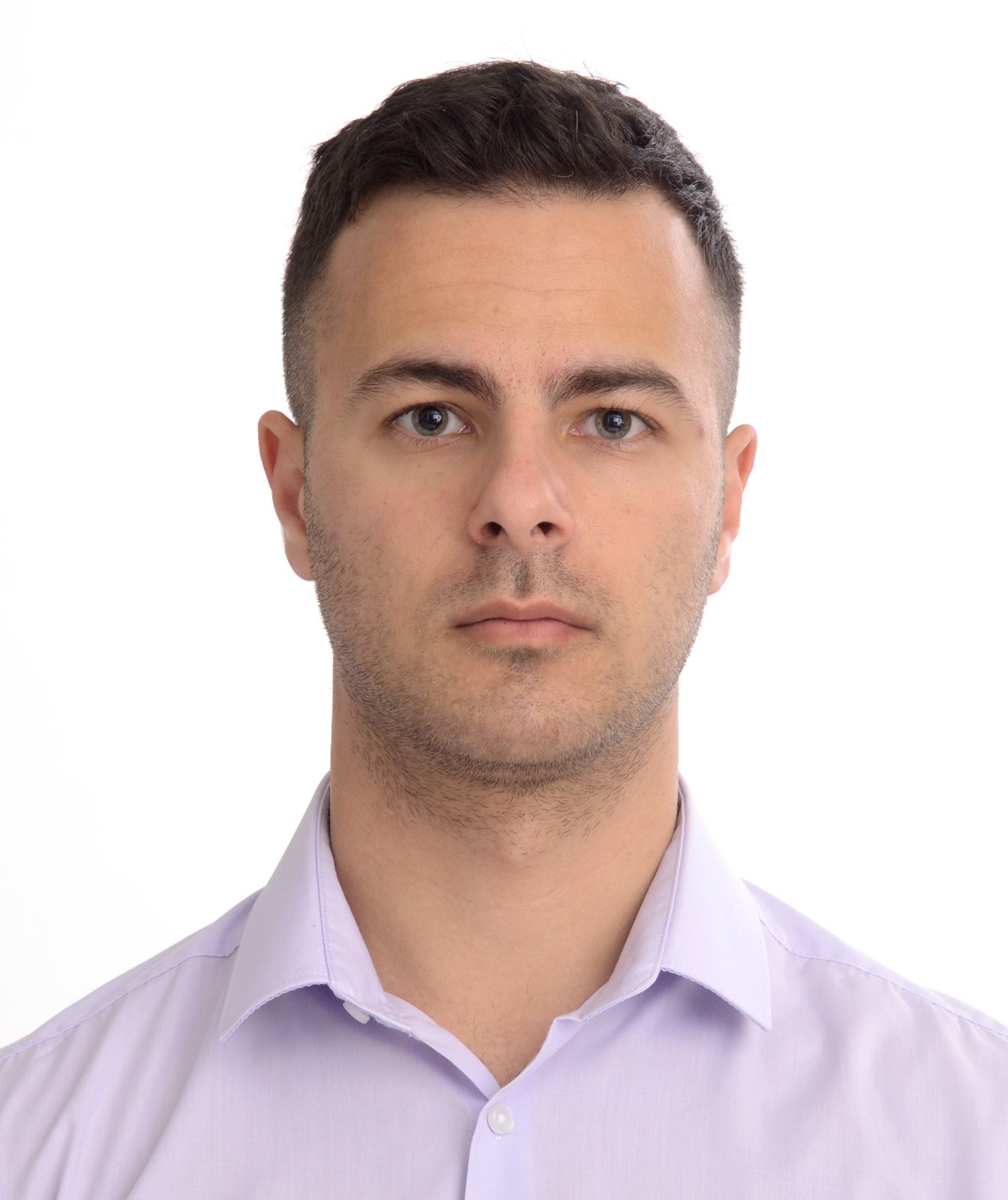}}]
{ANDREA LANUBILE } was born in Lecce, Italy. He received the B.Sc. and M.Sc. degrees in Automation and Control Engineering from the Politecnico di Milano, Italy. Since April 2024, he has been pursuing the Ph.D. degree at the Stanford Energy Control Lab in Stanford University. 

His research interests include state-of-health estimation of lithium-ion batteries, with a particular focus on identifying and tracking heterogeneities in battery packs.
\end{IEEEbiography}

\begin{IEEEbiography}[{\includegraphics[width=1in,height=1.25in,clip,keepaspectratio]{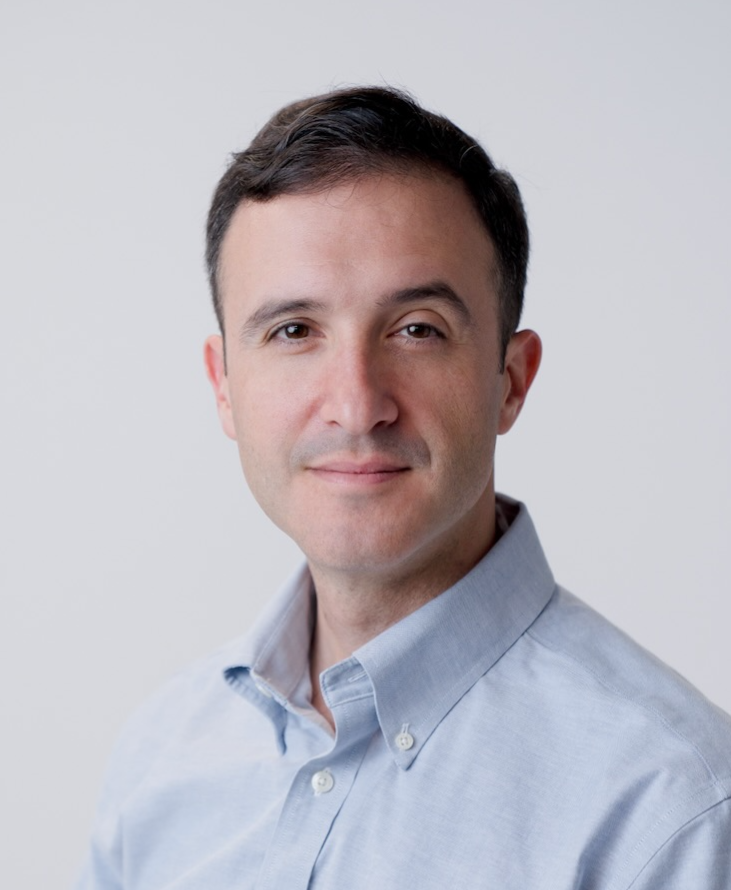}}]
{WILLIAM A. PAXTON } was born in New York, NY. He received the Ph.D. degree in Materials Science and Engineering from Rutgers University. He is Senior Staff Scientist and Project Lead at Volkswagen Group's Innovation Center California in Belmont, CA, where he leads research on electric-vehicle energy systems and advanced battery diagnostics. 

His work focuses on lithium-ion state-of-health modeling, second-life evaluation, and the integration of EV battery assets with grid and charging-infrastructure environments.
Paxton has authored publications spanning battery diagnostics, electrochemical materials, and electric-vehicle energy systems. Prior to his current role, he held electrification and materials-engineering positions at Ford Motor Company.
\end{IEEEbiography}

\begin{IEEEbiography}[{\includegraphics[width=1in,height=1.25in,clip,keepaspectratio]{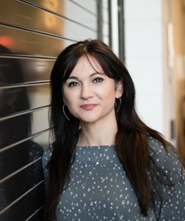}}]
{SIMONA ONORI } (Senior Member, IEEE) received her Ph.D. in Control Engineering from the University of Rome ``Tor Vergata'' in 2007. She is an Associate Professor of Energy Science and Engineering at Stanford University, where she leads the Stanford Energy Control Lab. 

Her research focuses on modeling, estimation, and optimization of electrochemical energy systems, including lithium-ion batteries, fuel cells, and hybrid electric powertrains. She has authored over 200 publications and holds several patents in energy storage and battery management. She is an SAE Fellow and Editor-in-Chief of the SAE International Journal of Electrified Vehicles, and has received several major awards.
\end{IEEEbiography}

\end{document}